\documentstyle[12pt,epsf,psfig,epsfig,twoside,amssymb,
rotating,colordvi,helvet,color,subfigure,axodraw]{article}

\advance\textheight by \topskip
\newcommand{\be}{\begin{equation}}
\newcommand{\ee}{\end{equation}}
\newcommand{\br}{\begin{eqnarray}}
\newcommand{\er}{\end{eqnarray}}
\newcommand{\ba}{\begin{array}}
\newcommand{\ea}{\end{array}}
\newcommand{\bi}{\begin{itemize}}
\newcommand{\ei}{\end{itemize}}
\newcommand{\bn}{\begin{enumerate}}
\newcommand{\en}{\end{enumerate}}
\newcommand{\bc}{\begin{center}}
\newcommand{\ec}{\end{center}}

\def\gsim{\buildrel{\scriptscriptstyle >}\over{\scriptscriptstyle\sim}}

\begin{document}

\begin{flushright}
{SHEP-06-04}\\
\today
\end{flushright}

\begin{center}
{\Large\bf Weak corrections to gluon-induced\\[0.15cm]
 top-antitop hadro-production}
\vskip1.0cm
{\large S. Moretti, M.R. Nolten and D.A. Ross}\\
\vskip0.5cm
{\it School of Physics \& Astronomy, University of Southampton,
Southampton SO17 1BJ, UK} 
\end{center}

\centerline{\bf Abstract}
\vskip0.5cm\noindent
We calculate purely weak virtual one-loop 
corrections to the production cross section
of top-antitop pairs at the Large Hadron Collider via the gluon-gluon 
fusion subprocess. We find very small
negative corrections to the total cross section, of order $-0.6\%$,
but significantly larger effects to the differential one, particularly 
in the transverse momentum distribution, of order 
$-5\%$ to $-10\%$ (in observable regions). 
In case of parity-conserving spin-asymmetries of the final state,
$\alpha_{\mathrm{S}}^2\alpha_{\mathrm{W}}$ corrections are typically
of a few negative percent, with the exception of positive and negative
peaks at $+12\%$ and $-5\%$, respectively (near where the tree-level 
predictions change sign), while  those arising in parity-violating
asymmetries (which are identically zero in QCD) are typically
at a level of a few permille. 

\section{Introduction}
\noindent
Top quark 
physics may well be the only context where both accurate Standard Model
(SM) tests and searches for new physics Beyond the Standard Model (BSM) will
be carried out at the Large Hadron Collider (LHC). If no BSM physics exists
at the TeV scale or the typical mass scale of new particles is (just) 
above the energy reach of the machine, one may well conceive that most of the
experimental and theoretical efforts will concentrate in establishing
the true nature of the top quark, which in turn will also enable one
to constrain possible manifestations of new physics.  While top quarks have
been discovered and studied at the Tevatron, the reduced number of events
available there will only allow one for a percent level determination
of the top mass (currently, $m_t=172.7\pm2.9$ GeV). This precision will
be improved by over a factor of two at the CERN machine. Here, one
will also be able  to measure
the top-quark width and quantum numbers (i.e., the electric charge and isospin,
accessible through
its Electro-Weak (EW) couplings). 
While there is certainly scope to investigate the
EW couplings of top-quarks by resorting to events with radiated photons
and $Z$ bosons \cite{myfirst}, the $V-A$ structure (or otherwise) 
of the charged decay current
can already be probed directly in $t\bar t$ events, if one 
recalls that the top-(anti)quark decays into a 
bottom-(anti)quark and a $W$ boson rather than hadronising. 
Finally, for the same reason, 
the top-(anti)quark transmits its spin properties to the
decay products rather efficiently, so that the latter can be explored 
in suitable experimental observables \cite{ttpol,review,Marina}. 

Clearly, in order to perform all the relevant 
measurements in $t\bar t$ events, any source of SM corrections
should be well under control. While complete one-loop results exist for QCD 
\cite{ttNLO}, similar weak effects have been unavailable until very recently 
\cite{Hans,Werner}. These last two papers were concerned with purely weak 
$\alpha_{\mathrm{S}}^2\alpha_{\mathrm{W}}$  effects entering  
the $q\bar q\to t\bar t$ subprocess only. It is the purpose of this paper
to complement those studies, by computing the
corrections  to the $gg\to t\bar t$ channel, which is
in fact dominant at the LHC (whereas the quark initiated one is the
leading partonic component at the Tevatron). Early, though incomplete results,
 for $\alpha_{\mathrm{S}}^2\alpha_{\mathrm{W}}$  corrections to top-antitop
hadro-production can be found in Ref.~\cite{earlycalc} for both
$gg$ and $q\bar q$ initiated subprocesses. Concerning the $gg\to t\bar t$ 
case, unlike Ref.~\cite{earlycalc},
notice that we have also included here the one-loop triangle contributions 
for $gg\to Z^*\to t\bar t$, which are in fact non-zero for off-shell
$Z$ bosons. (Recall that the Landau-Yang's theorem \cite{LandauYang}
is only valid for on-shell $Z$ bosons and we have explicitly verified 
this to be the case in our calculation if we take the appropriate 
limit.) We have also updated the analyses of \cite{earlycalc} to the
most recent top and Higgs mass values as well as Parton Distribution
Functions (PDFs).

The paper is organised as follows. The next section illustrates the importance
that EW effects should have at TeV energy scales. Sections 3 and 4 
will be devoted to describe our computation and present the numerical results,
respectively. The last section contains our conclusions.

\section{EW effects at TeV scale energies}
\noindent
The purely weak (W) component of Next-to-Leading Order (NLO)  
EW effects produces 
corrections of the type $\alpha_{\rm{W}}\log^2({\mu^2}/M_W^2)$, where 
$\alpha_{\mathrm{W}}\equiv \alpha_{\mathrm{EM}}/\sin^2\theta_W$,
with $\alpha_{\mathrm{EM}}$ the Electro-Magnetic (EM) coupling
constant and $\theta_W$ the Weinberg angle. Here, $\mu$ represents
some typical energy scale affecting the top-antitop process 
in a given observable, e.g., the transverse momentum of either the top
(anti)quark or the top-antitop invariant mass. 
For large enough $\mu$ values,  such EW effects may be competitive not
only with Next-to-NLO (NNLO) (as $ \alpha_{\rm{W}}\approx 
\alpha_{\rm{S}}^2$) but also with NLO QCD corrections (e.g., for
${\mu}=0.5$ TeV, $\log^2({\mu^2}/M_W^2)\approx10$).

These `double logs' are of Sudakov origin and are  
due to a lack of cancellation between virtual and real $W$-emission in
higher order contributions. This is in turn a consequence of the 
violation of the Bloch-Nordsieck theorem in non-Abelian theories
\cite{Denner-Pozzorini,Ciafaloni:2000df-Ciafaloni:2000rp}.
The problem is in principle present also in QCD. In practice, however, 
it has no observable consequences, because of the final averaging of the 
colour degrees of freedom of partons, forced by confinement
into colourless hadrons. This does not occur in the EW case,
where the initial state generally has a non-Abelian charge,
as in proton-proton scattering. Besides, these
logarithmic corrections are finite (unlike in
QCD), since $M_W$ provides a physical
cut-off for $W$-emission. Hence, for typical experimental
resolutions, softly and collinearly emitted weak bosons need not be included
in the production cross section and one can restrict oneself to the 
calculation
of weak effects originating from virtual corrections only. 
By doing so, similar
logarithmic effects, $\sim\alpha_{\rm{W}}\log^2({\mu^2}/M_Z^2)$, 
are generated also by $Z$-boson corrections.
Finally, in some instances all these purely weak contributions can  be
isolated in a gauge-invariant manner from EM effects which therefore may not
be included in the calculation (as it is the case here). (Besides,
EM corrections are not subject to Sudakov enhancement.) 
In view of all this,  it becomes of crucial importance to assess
quantitatively such weak corrections
affecting, in particular, a key process 
(for both present and future 
hadron colliders) such as top-antitop hadro-production. 
\vspace*{-0.15truecm}
\section{Calculation}
\noindent
It is the aim of our paper to report on the computation of the full
one-loop weak effects entering the subprocess $gg\to t\bar t$, through the
perturbative order $\alpha_{\mathrm{S}}^2\alpha_{\mathrm{W}}$. 
We will instead ignore altogether the contributions
of tree-level $\alpha_{\mathrm{S}}^2\alpha_{\mathrm{W}}$ terms
involving the radiation (bremsstrahlung)
of real $Z$ bosons. 
In the Feynman-'t Hooft   gauge\footnote{Here,
we mean that the numerator of the massive 
gauge boson propagators is taken to be $-ig_{\mu\nu}$ and Goldstone
 bosons, with masses equal to their gauge boson counterparts, are
included where appropriate.}, the one used for this calculation,
neglecting the $b$-mass, one has to calculate the
following one-loop prototype diagrams 
for $gg\to t\bar t$, ignoring permutations of external gluons:

\begin{minipage}{\textwidth}
\vspace{-0.5cm}\hspace{-0.5cm}
\includegraphics[width=.25\linewidth]{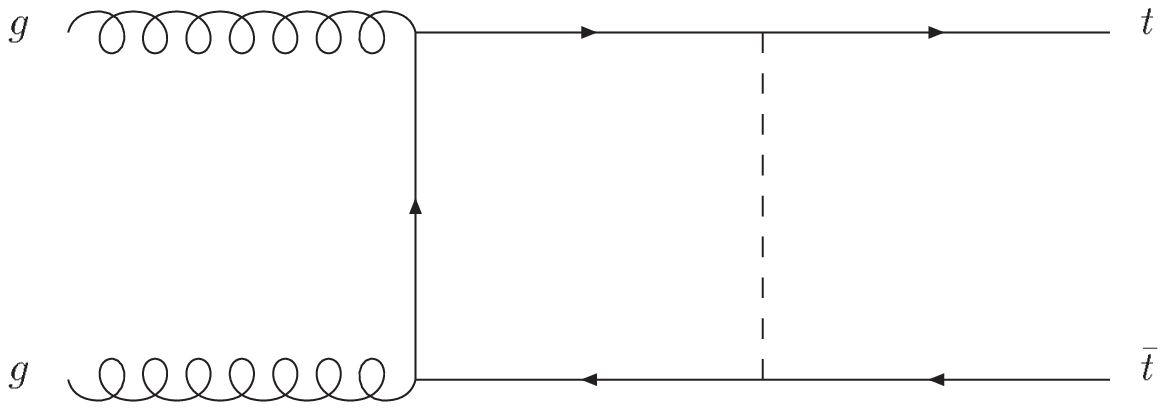}
\includegraphics[width=.25\linewidth]{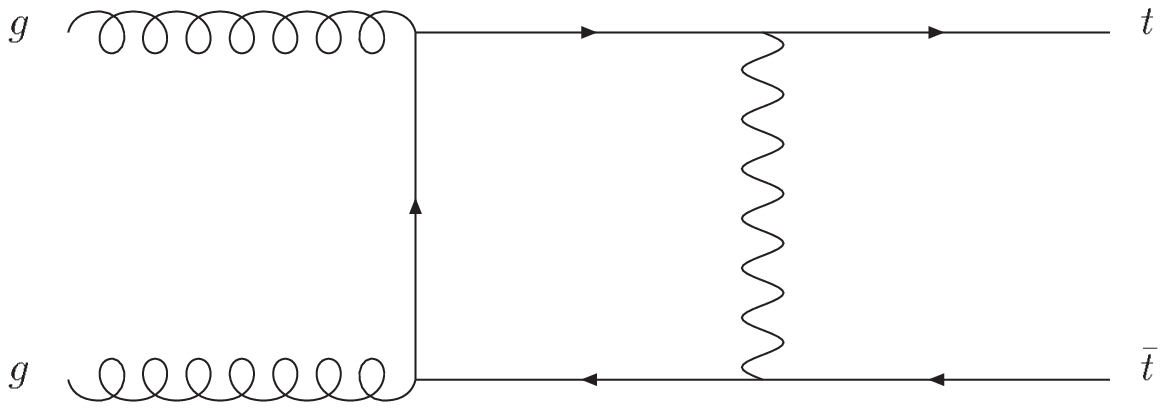}
\end{minipage}

\begin{minipage}{\textwidth}
\vspace{-4.0cm}\hspace{-0.5cm}
\includegraphics[width=.25\linewidth]{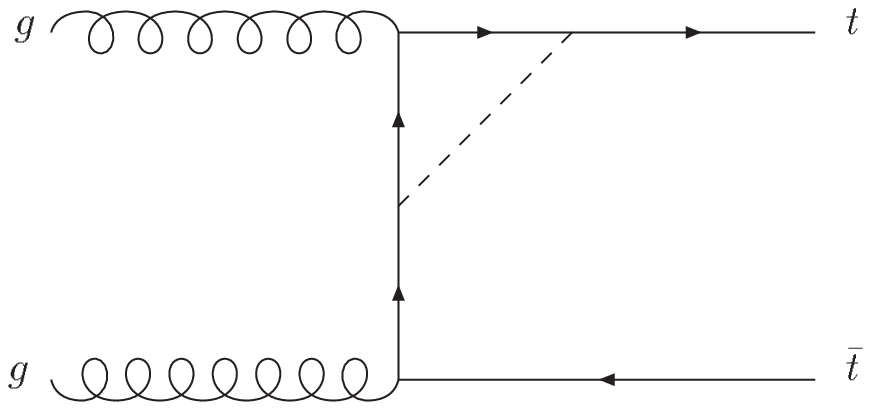}
\includegraphics[width=.25\linewidth]{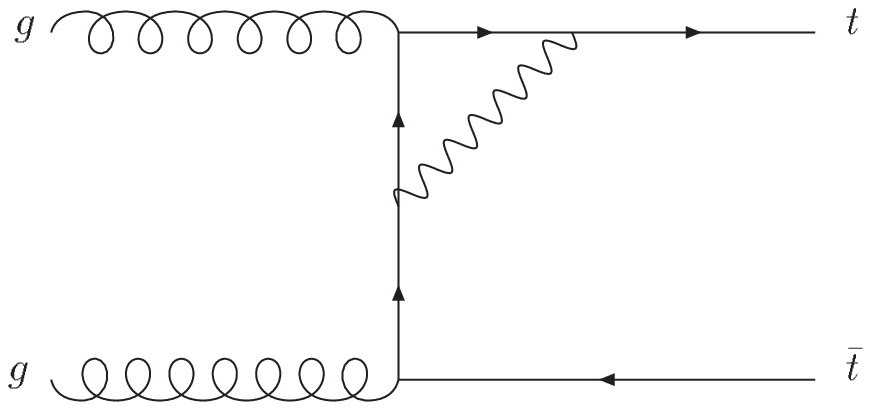}
\end{minipage}

\begin{minipage}{\textwidth}
\vspace{-4.0cm}\hspace{-0.5cm}
\includegraphics[width=.25\linewidth]{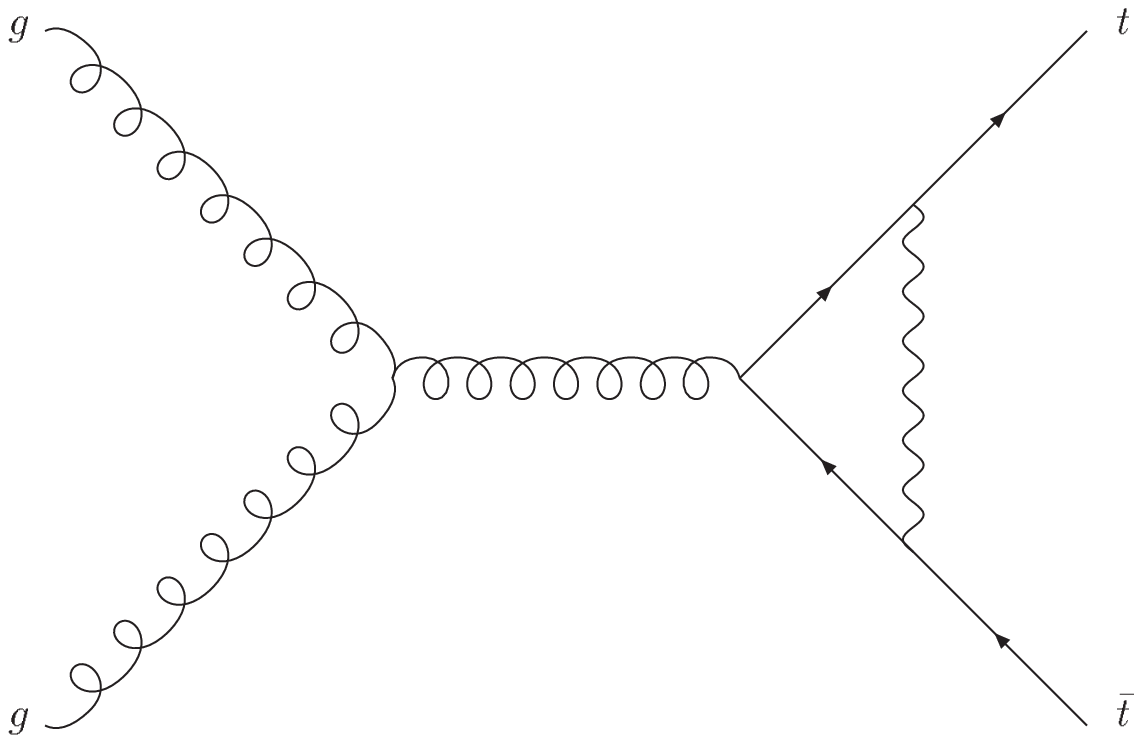}
\includegraphics[width=.25\linewidth]{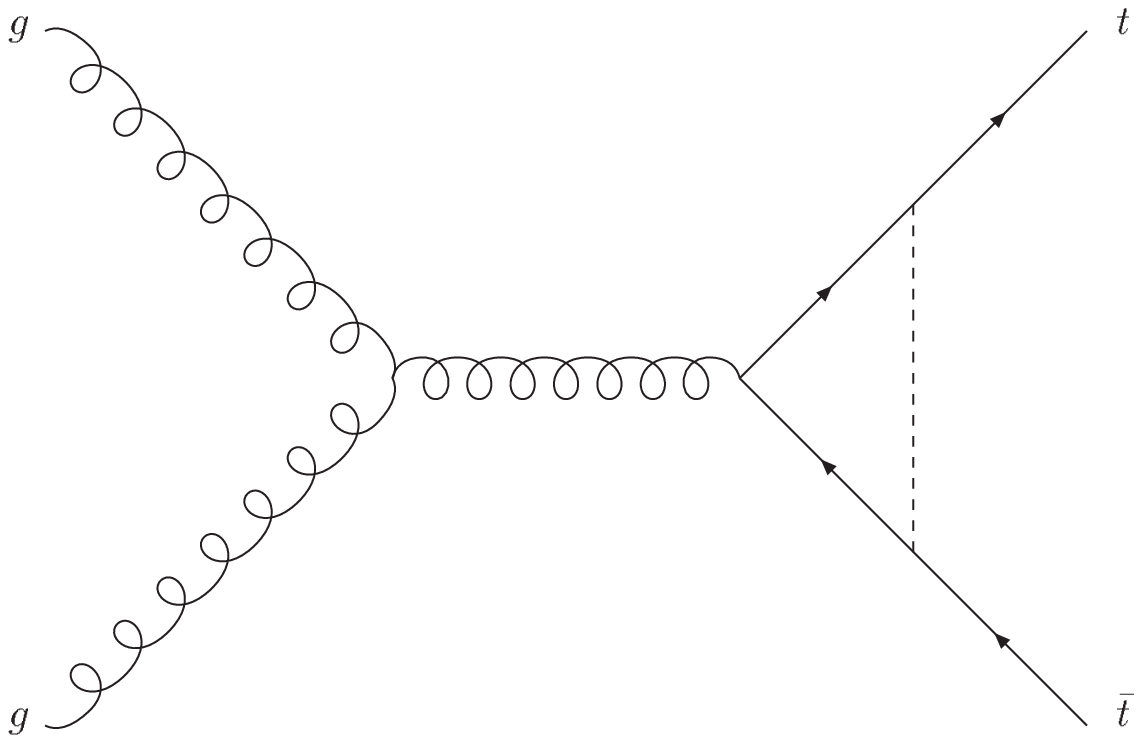}
\includegraphics[width=.25\linewidth]{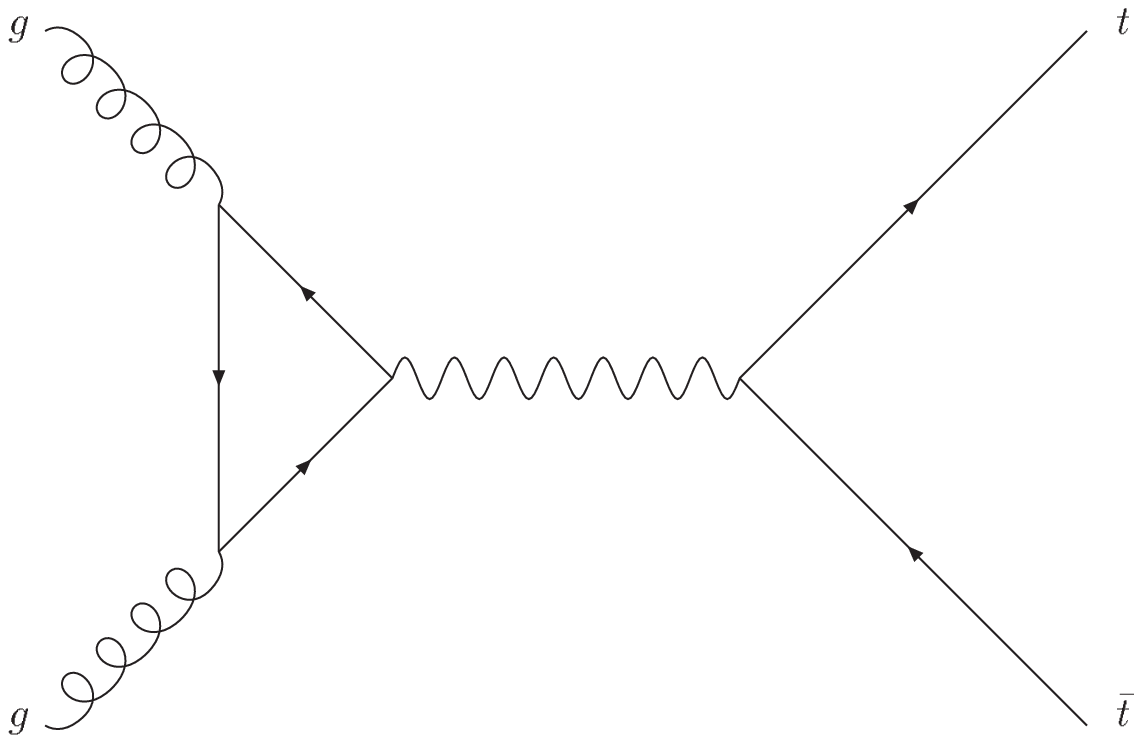}
\includegraphics[width=.25\linewidth]{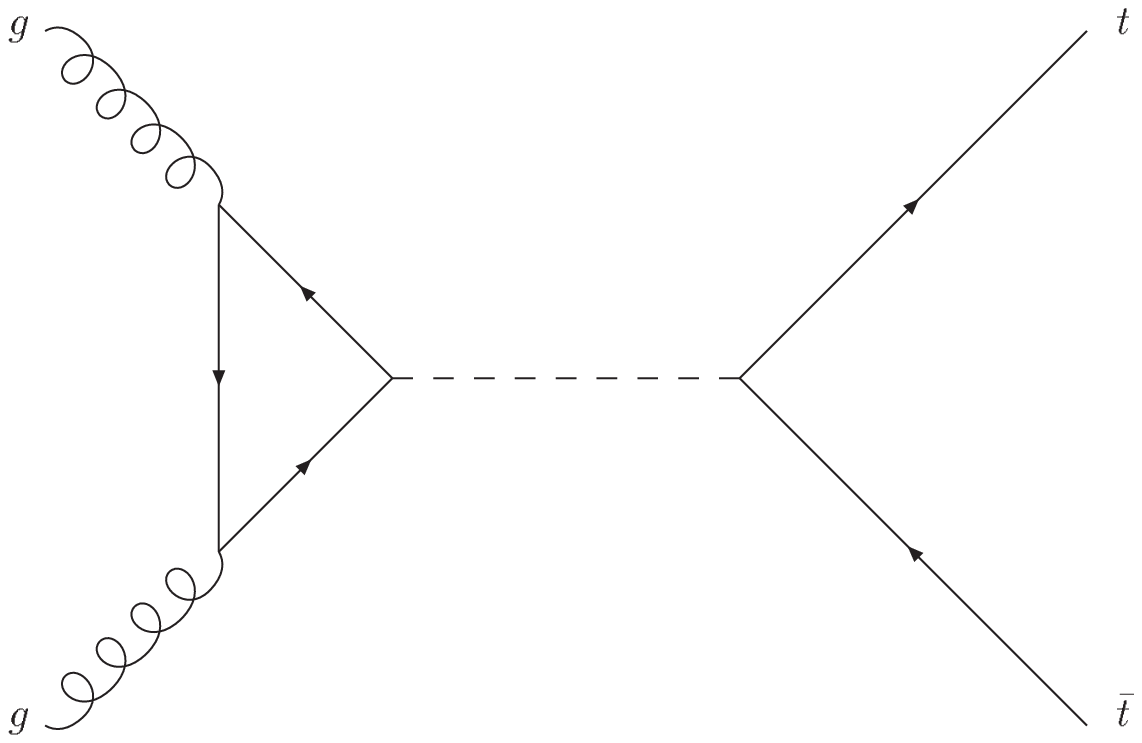}
\end{minipage}

\begin{minipage}{\textwidth}
\vspace{-0.5cm}\hspace{-0.5cm}
\includegraphics[width=.25\linewidth]{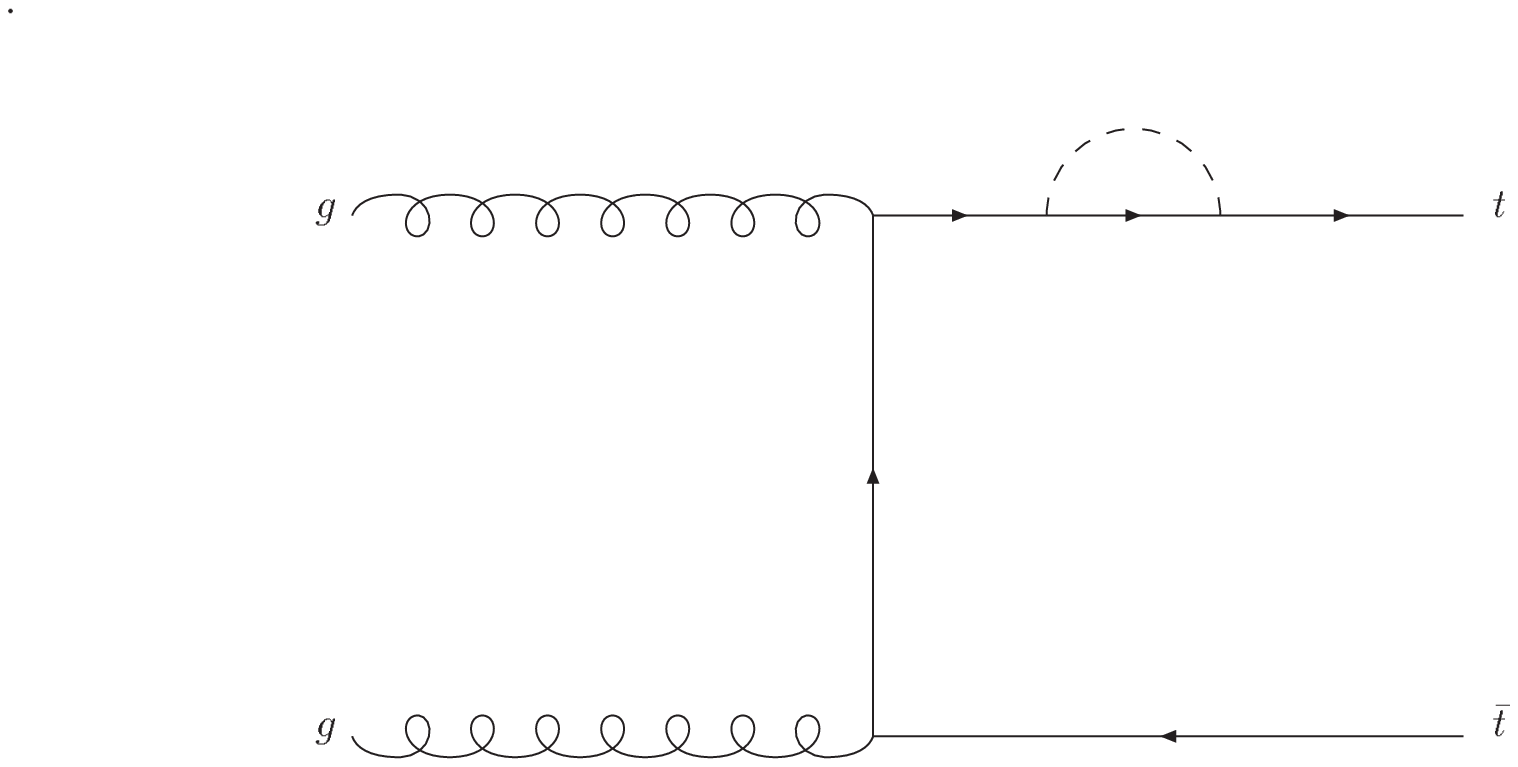}
\includegraphics[width=.25\linewidth]{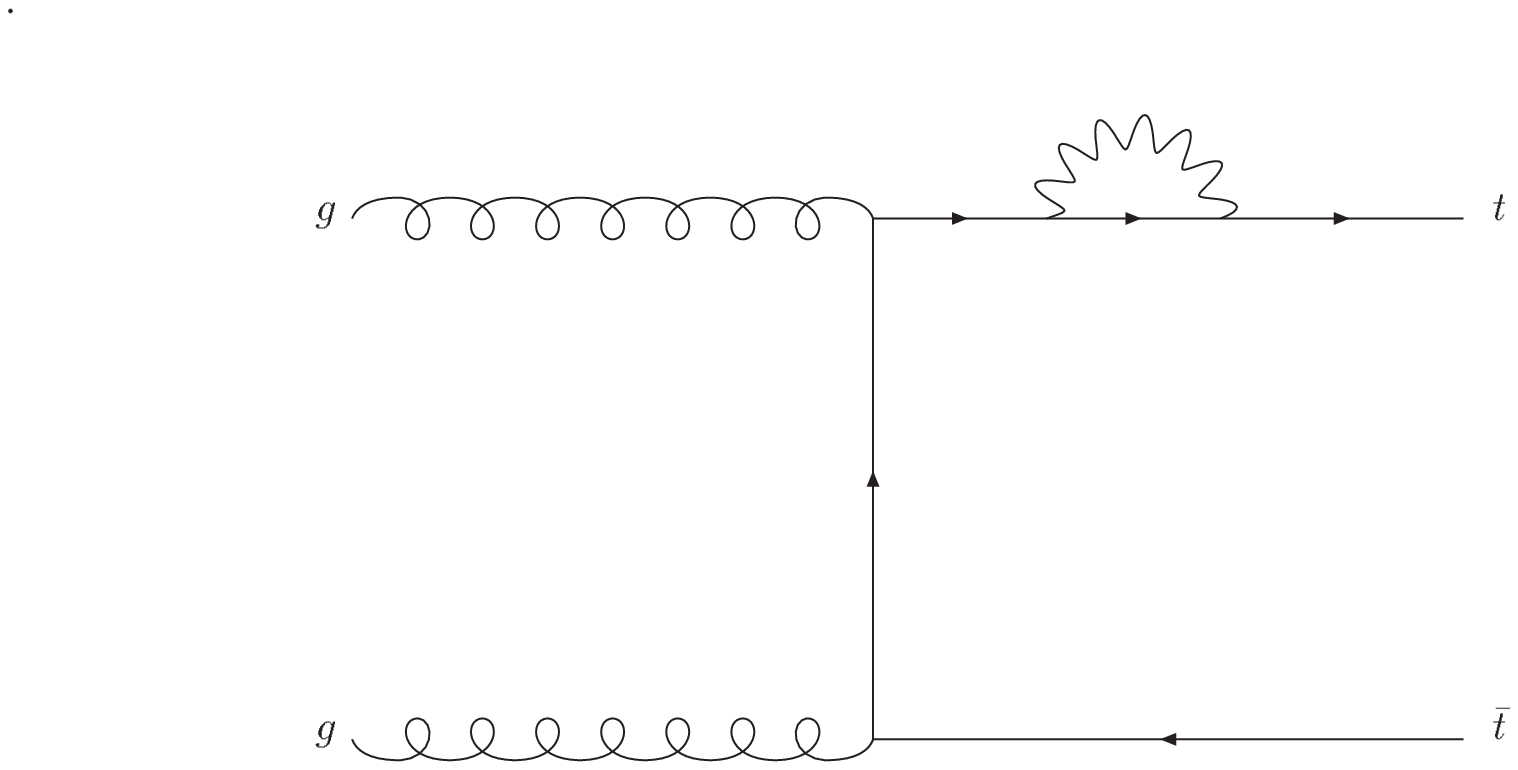}
\end{minipage}

\begin{minipage}{\textwidth}
\vspace{-3.5cm}\hspace{-0.5cm}
\includegraphics[width=.25\linewidth]{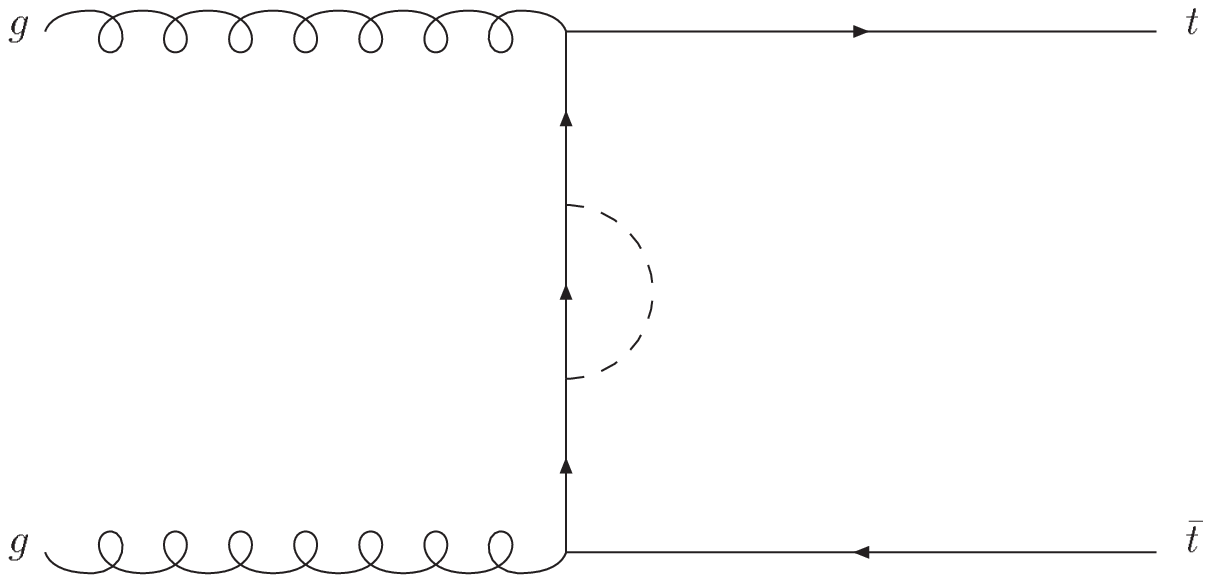}
\includegraphics[width=.25\linewidth]{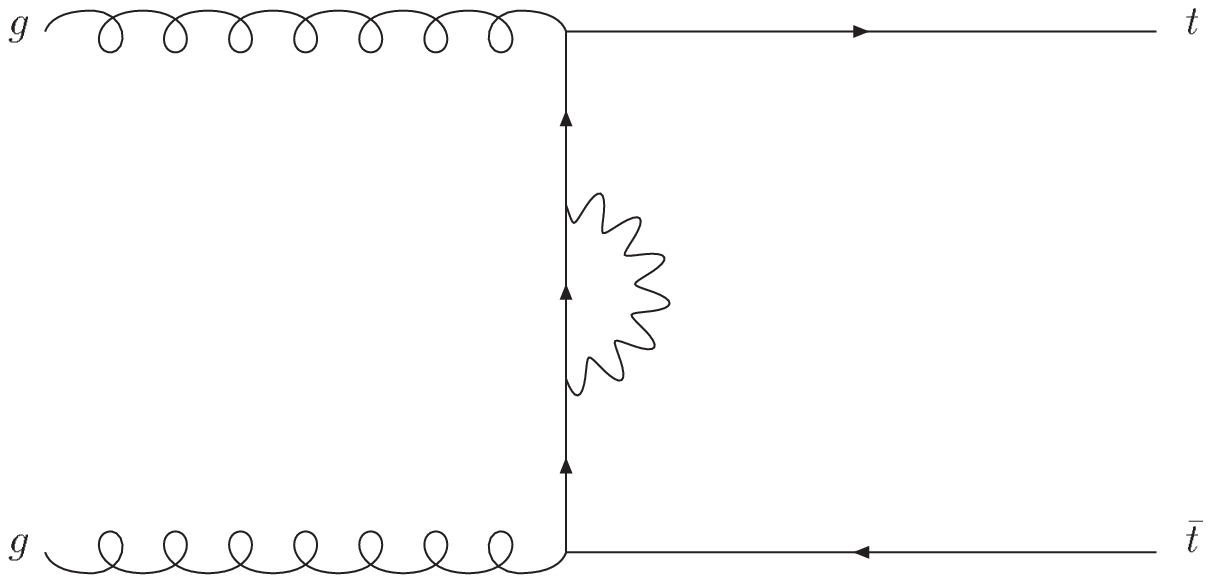}
\end{minipage}

\begin{minipage}{\textwidth}
\vspace{-3.5cm}\hspace{-0.5cm}
\includegraphics[width=.25\linewidth]{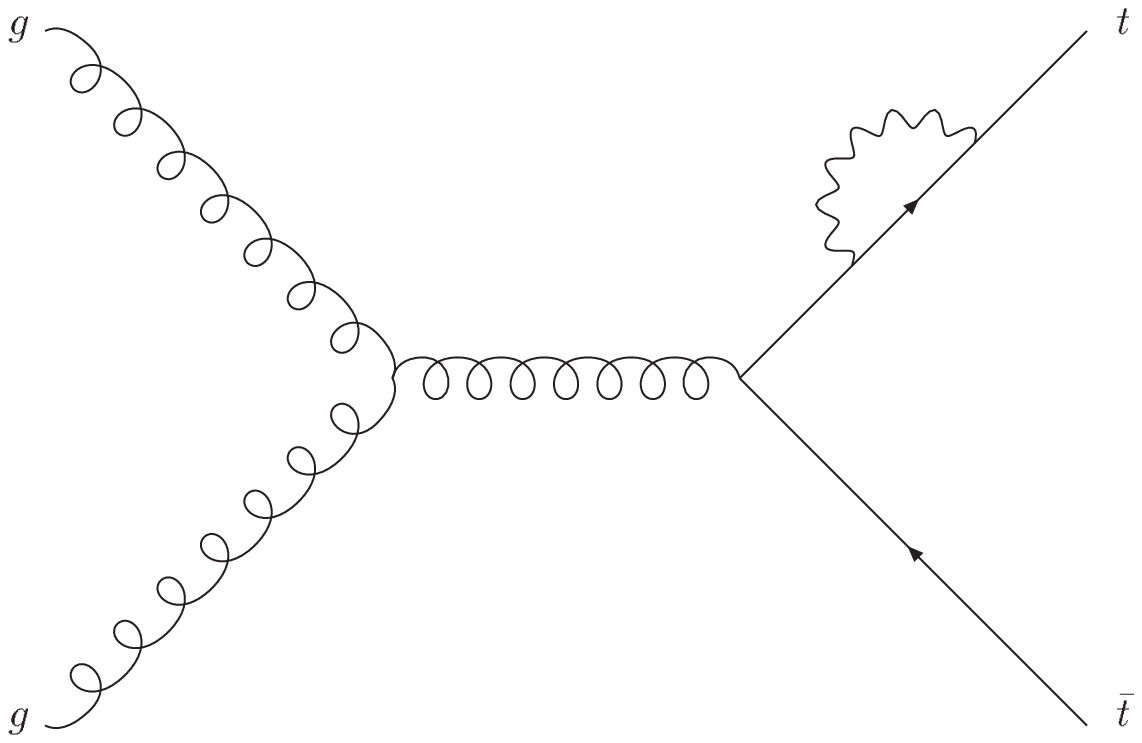}
\includegraphics[width=.25\linewidth]{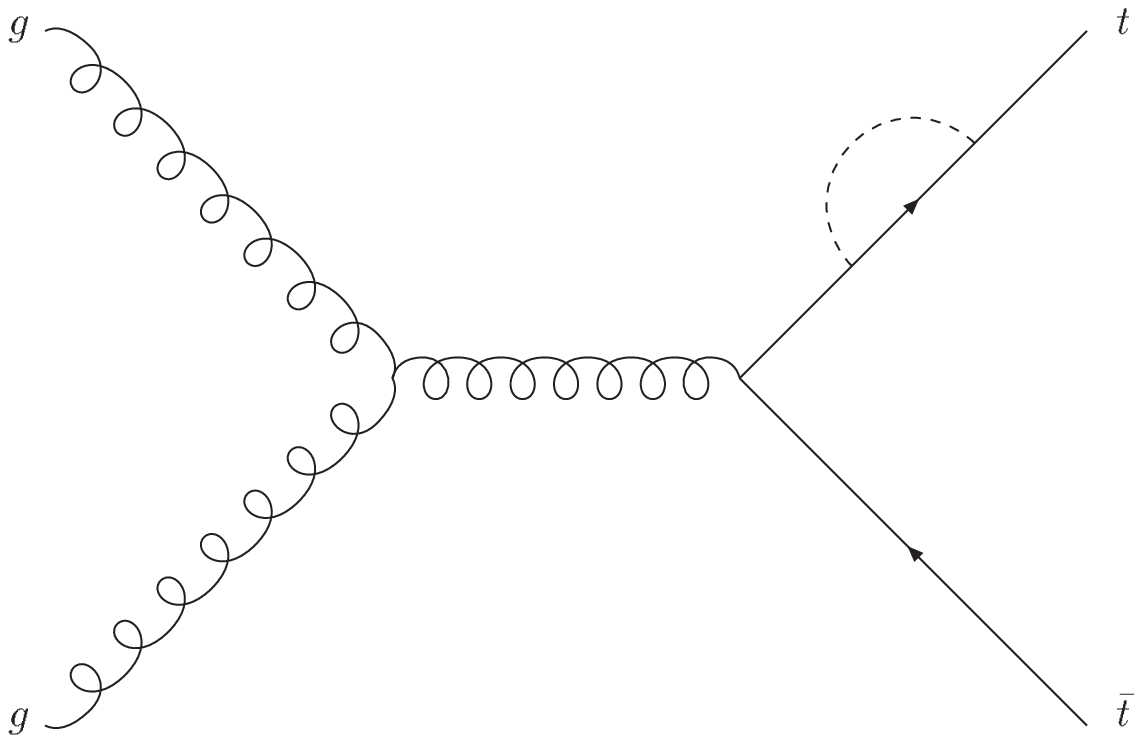}
\end{minipage}

(In the calculation of all such graphs, we have retained the full spin 
dependence of the final state particles.)

Given the large number of 
diagrams involved in the computation,
it is of paramount
importance to perform careful checks. In this respect, we 
should mention that our  expressions 
have been calculated independently
by at least two of us using FORM \cite{Vermaseren:2000nd} and 
that some results have also been reproduced by another
program based on FeynCalc \cite{Kublbeck:1990xc}. 
Upon removing the one-loop triangle contributions 
for $gg\to Z^*\to t\bar t$ and for identical choices of Higgs mass, we
also reproduce well the results of the first paper in Ref.~\cite{earlycalc}. 
(In fact, this triangle 
contribution is always only marginally relevant.) Finally,
we find reasonable agreement with 
Ref.~\cite{Marina} (see also \cite{Claudio,newClaudio}) in the Sudakov limit, 
i.e.,
for large invariant masses and transverse momenta of the final state,
provided that the final state particles are fairly central (see later on).

Some of the diagrams  contain ultraviolet divergences.
In the case of self-energies for massive particles the mass subtraction
has been effected on mass-shell so that the masses refer to the
physical (pole) masses. The remaining divergences
 have been subtracted using the `modified' Dimensional Reduction
(${\overline{\mathrm{DR}}}$) scheme at
the scale $\mu=M_Z$. The use of ${\overline{\mathrm{DR}}}$,
as opposed to the more usual `modified' Minimal Subtraction
(${\overline{\mathrm{MS}}}$) scheme, is forced
upon us by the fact that the $W$ and $Z$ bosons contain axial couplings
which cannot be consistently treated in ordinary dimensional 
regularisation. The strong coupling is {\sl not} renormalised 
by the weak interactions, which means that there are Ward identities
which cancel the divergent corrections to
the strong coupling. Thus the choice of subtraction scheme
has no effect on our final results since the scheme dependence cancels when 
all graphs are summed over. On the other hand the the EM coupling,
$\alpha_{\mathrm{EM}}$, 
has been taken to be $1/128$ in order to correctly account for the 
SM running of the electroweak coupling up to the threshold for
$t\bar{t}$ production.

For the top mass and width, the latter entering some of the loop diagrams,
 we have taken $m_t=175$ GeV and $\Gamma_t=1.55$ GeV, respectively. (As already
intimated, the $b$-quark was considered massless.)
The $Z$ mass used was $M_Z=91.19$ GeV and was related to the $W$ mass, $M_W$, via the
SM formula $M_W=M_Z\cos\theta_W$, where $\sin^2\theta_W=0.232$.
(Corresponding widths were $\Gamma_Z=2.5$ GeV and $\Gamma_W=2.08$ GeV.)
The Higgs boson mass and width were set to 150 GeV and 16 MeV by default,
respectively. However, other mass (and consequently width)
choices (above the LEP limit of $M_H\gsim115$ GeV)
have been investigated (see later on).
The PDFs we have used are CTEQ6L1
\cite{cteq6} taken at the factorisation scale $Q=2m_t$. We have also checked
other sets, but found no significant difference in the relative size of our
corrections.

\section{Numerical results}

Our initial findings are presented in Figs.~1--2. Here, we consider both 
differential spectra of some kinematic observables as well as global
asymmetries, the latter plotted, e.g., against the invariant mass
of the $t\bar t$ pair. Notice that this last quantity
can only be defined when both the top
and anti-top four-momenta are reconstructed, which happens in the
case of fully hadronic and semi-leptonic/hadronic decays\footnote{In the
second case one would reconstruct the longitudinal neutrino momentum by
equating the transverse one to the missing transverse energy and enforcing
the $W$ mass reconstruction.}, but not for fully leptonic ones 
(where two neutrinos escape detection).

The definitions of the asymmetries are as follows:
\begin{eqnarray}\label{asymmetries}                                \nonumber
A_{LL} \, d\sigma &\equiv &\,d\sigma_{++}\,   - \, d\sigma_{+-}  
                   +       \,d\sigma_{--}\,   - \, d\sigma_{-+},\\ \nonumber
~~A_{L} \,  d\sigma &\equiv &\,d\sigma_{- }\, ~~- \, d\sigma_{+ },\\ 
A_{PV} \, d\sigma &\equiv &\,d\sigma_{--}\,   - \, d\sigma_{++}.
\end{eqnarray}
For $A_L$ only the polarisation of {\sl either} the $t$-quark or 
${t}$-antiquark is assumed to be measured, whereas  the other two asymmetries
require the determination of the polarisations
 of {\sl both} the outgoing particles. 
$A_{LL}$ is parity-conserving while the other two are 
parity-violating\footnote{See Ref.~\cite{ttpol} for a choice of
observables correlated to these asymmetries.}.
Here, the indices $+$ and $-$ refer to the helicities of right (R) and left (L)
handed (anti)top quark, respectively.
(Other basis choices are also possible, see Ref.~\cite{ttpol}.)
 
We find that the overall effect of our 
${\cal O}(\alpha_{\mathrm{S}}^2\alpha_{\mathrm{W}})$ corrections is
about $-0.6\%$ at the inclusive level (i.e., to the total
cross section, as obtained
from the integral of any of the curves in Fig.~1). However, 
for differential cross sections, 
effects can be of either sign, notably in the (anti)top transverse
momentum and top-antitop invariant mass. 
For an LHC integrated luminosity of
300 fb$^{-1}$, differential rates of order a few 10$^{-5}$ pb may yield
detectable events (after accounting for decay fractions, 
tagging efficiency and 
reconstruction performance).  In the corresponding observable 
kinematic range, the maximum correction occurs
for the transverse momentum 
spectrum of the (anti)top at around 1 TeV  (in the Sudakov limit), 
where it reaches almost
the $-10\%$ level. In the same kinematic regime, the effects are
smaller for  the invariant mass and pseudorapidity distributions. At small
transverse momentum, invariant mass as well as in the very 
forward/backward direction is  where the corrections are positive,
with a maximum of ${\cal O}(+2\%)$ for the second of these observables.

In the case of the parity-conserving asymmetry, for which the
${\cal O}(\alpha_{\mathrm{S}}^2)$ result is non-zero, 
${\cal O}(\alpha_{\mathrm{S}}^2\alpha_{\mathrm{W}})$ effects enter
significantly (up to the +12 and $-5\%$ level or so)
only near the point where tree-level predictions are zero. Otherwise, they
amount to a few negative percent at the most. For the parity-violating 
asymmetries, the
relevant quantity is the actual value of the 
${\cal O}(\alpha_{\mathrm{S}}^2\alpha_{\mathrm{W}})$ result, as the
${\cal O}(\alpha_{\mathrm{S}}^2)$ term is identically zero. In both cases,
the rates are at the permille level (away from the $M_{t\bar t}\approx2m_t$
threshold, where our fixed order results are not fully reliable).

\begin{figure}[!t]
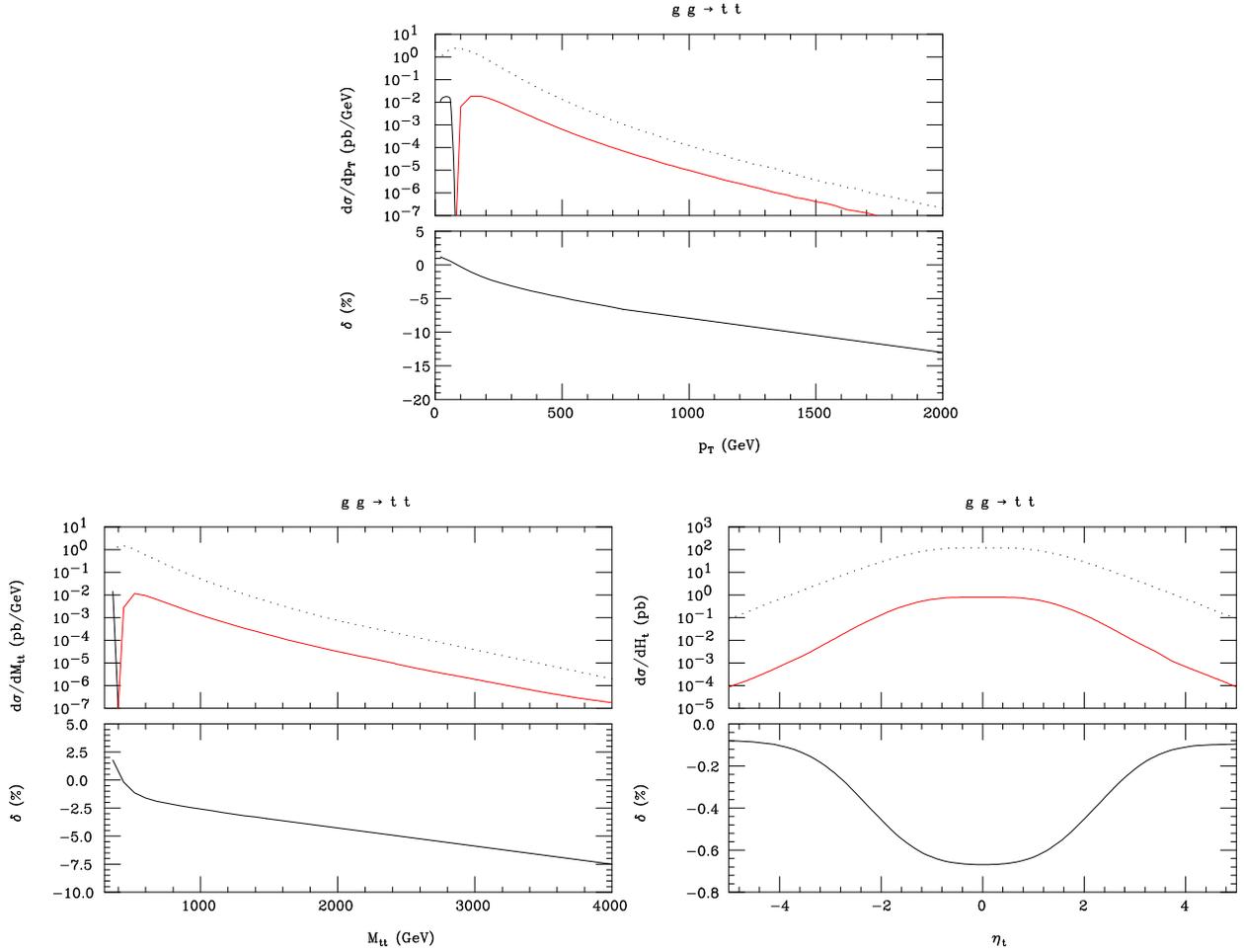

\begin{minipage}{\textwidth}
\hspace{3.95cm}
\includegraphics[width=.3725\linewidth,angle=90]{pT.ps}
\end{minipage}
\begin{minipage}{\textwidth}
\vspace{0.5cm}\hspace{-0.5cm}
\includegraphics[width=.3725\linewidth,angle=90]{M.ps}
\includegraphics[width=.3725\linewidth,angle=90]{eta.ps}
\caption{Differential distributions of the subprocess
$gg\to t\bar t$ through the ${\cal O}(\alpha_{\mathrm{S}}^2)$
(top frames, dotted) and the 
${\cal O}(\alpha_{\mathrm{S}}^2\alpha_{\mathrm{W}})$ 
(top frames, solid) 
as well as the percentage of the latter with respect to the former
(bottom frames, solid)  
for the (anti)top transverse momentum $p_T$, the top-antitop invariant mass
$M_{t\bar t}$ and the (anti)top pseudorapidity $\eta_t$. 
(Lightly/Red coloured solid tracts in logarithmic scale 
are intended to be negative.)}
\end{minipage}
\end{figure}

\begin{figure}[!t]
\begin{minipage}{\textwidth}
\hspace{3.95cm}
\includegraphics[width=.3725\linewidth,angle=90]{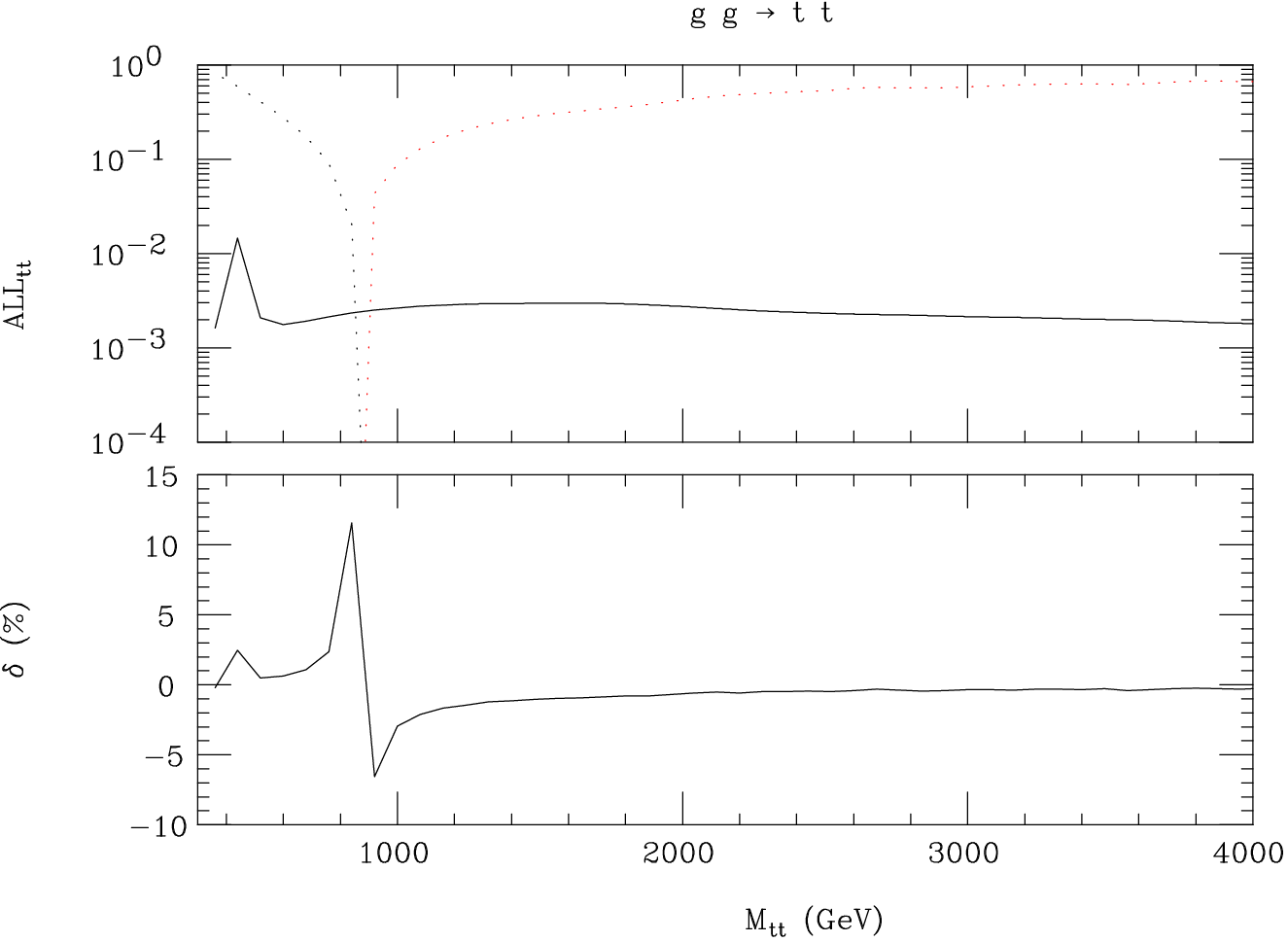}
\end{minipage}
\begin{minipage}{\textwidth}
\vspace{0.5cm}\hspace{-0.5cm}
\includegraphics[width=.3725\linewidth,angle=90]{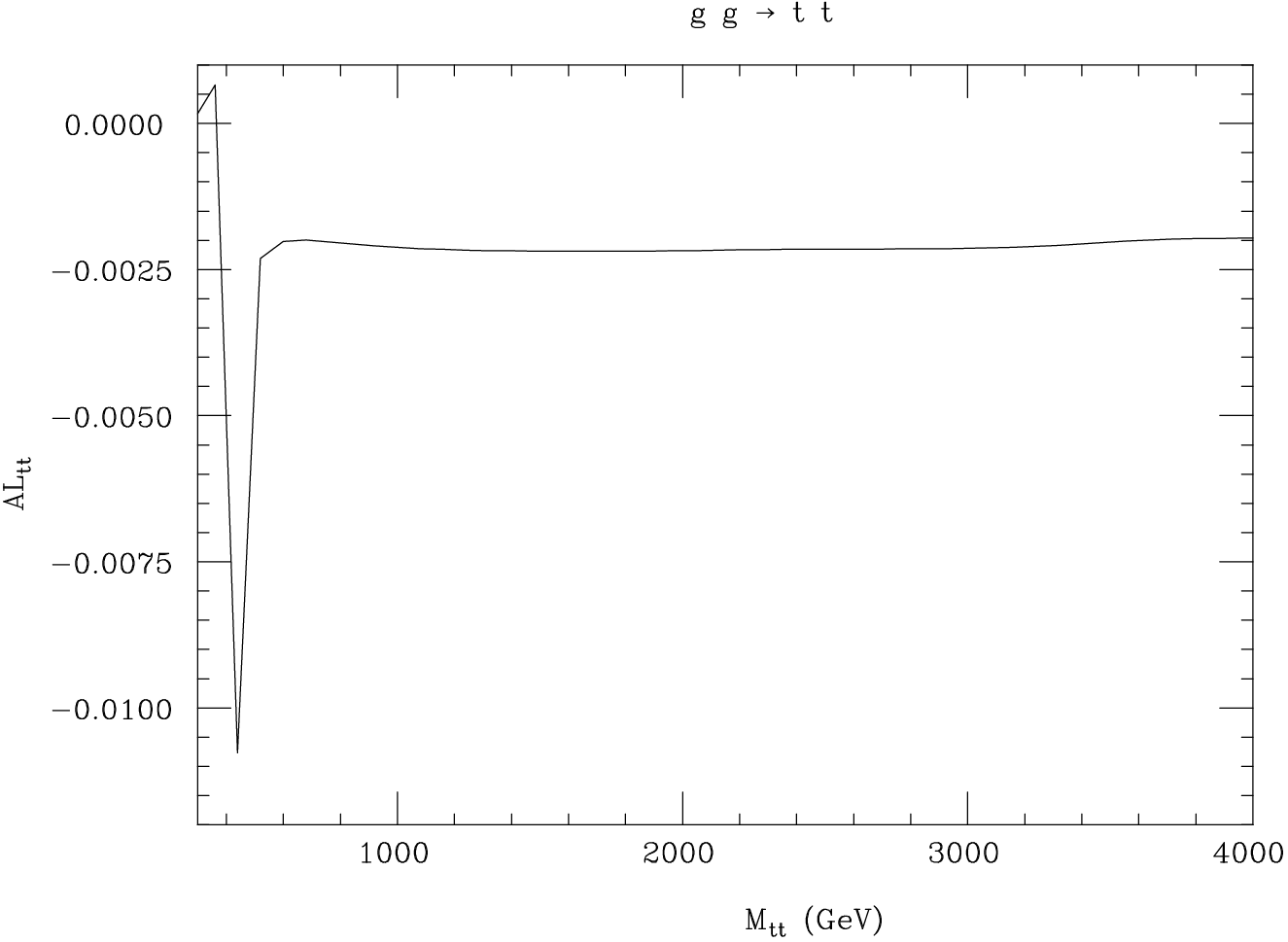}
\includegraphics[width=.3725\linewidth,angle=90]{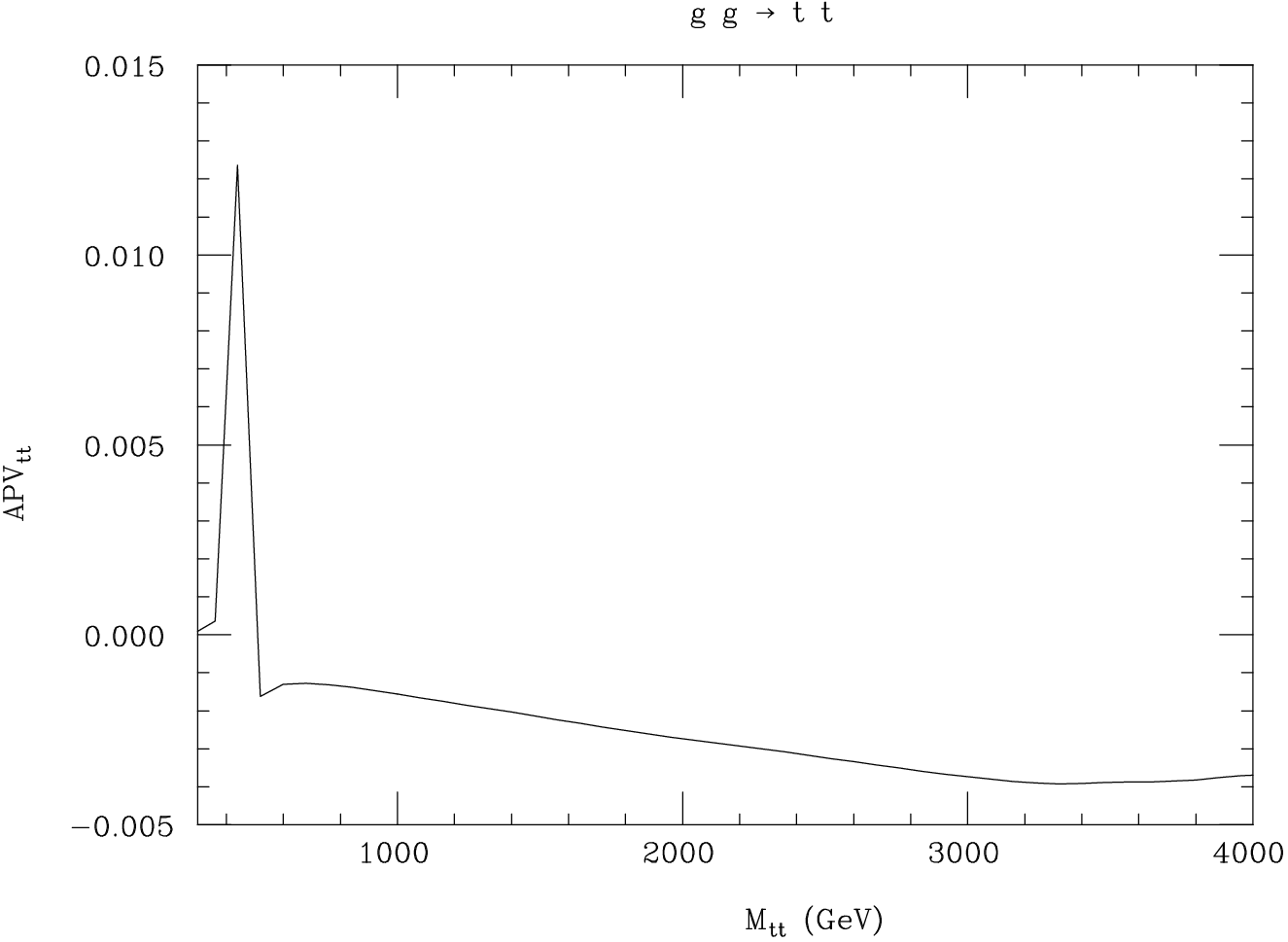}
\caption{The differential spin asymmetry $A_{LL}$ (as defined in the text)
of the subprocess
$gg\to t\bar t$ through the ${\cal O}(\alpha_{\mathrm{S}}^2)$
(top frame, dotted) and the 
${\cal O}(\alpha_{\mathrm{S}}^2\alpha_{\mathrm{W}})$ 
(top frame, solid).
(Note that the LO QCD contribution changes sign
at $\approx900$ GeV and is heavily dependent on $M_{t\bar{t}}$ whereas the 
${\cal O}(\alpha_{\mathrm{S}}^2\alpha_{\mathrm{W}})$ 
correction is not.) Just below the top frame we show the 
percentage correction
to the (non-zero) LO QCD asymmetry for $A_{LL}$ due to 
${\cal O}(\alpha_{\mathrm{S}}^2\alpha_{\mathrm{W}})$  effects.
The  lower two frames display the asymmetries $A_L$ and $A_{PV}$
(as defined in the text), which vanish
exactly in LO QCD, through the same order.
The 
asymmetries are calculated along
the helicity axis as a function of the top-antitop invariant mass
$M_{t\bar t}$. 
}
\end{minipage}
\end{figure}

\begin{figure}[!t]
\begin{minipage}{\textwidth}
\hspace{-0.5cm}
\includegraphics[width=.375\linewidth,angle=90]{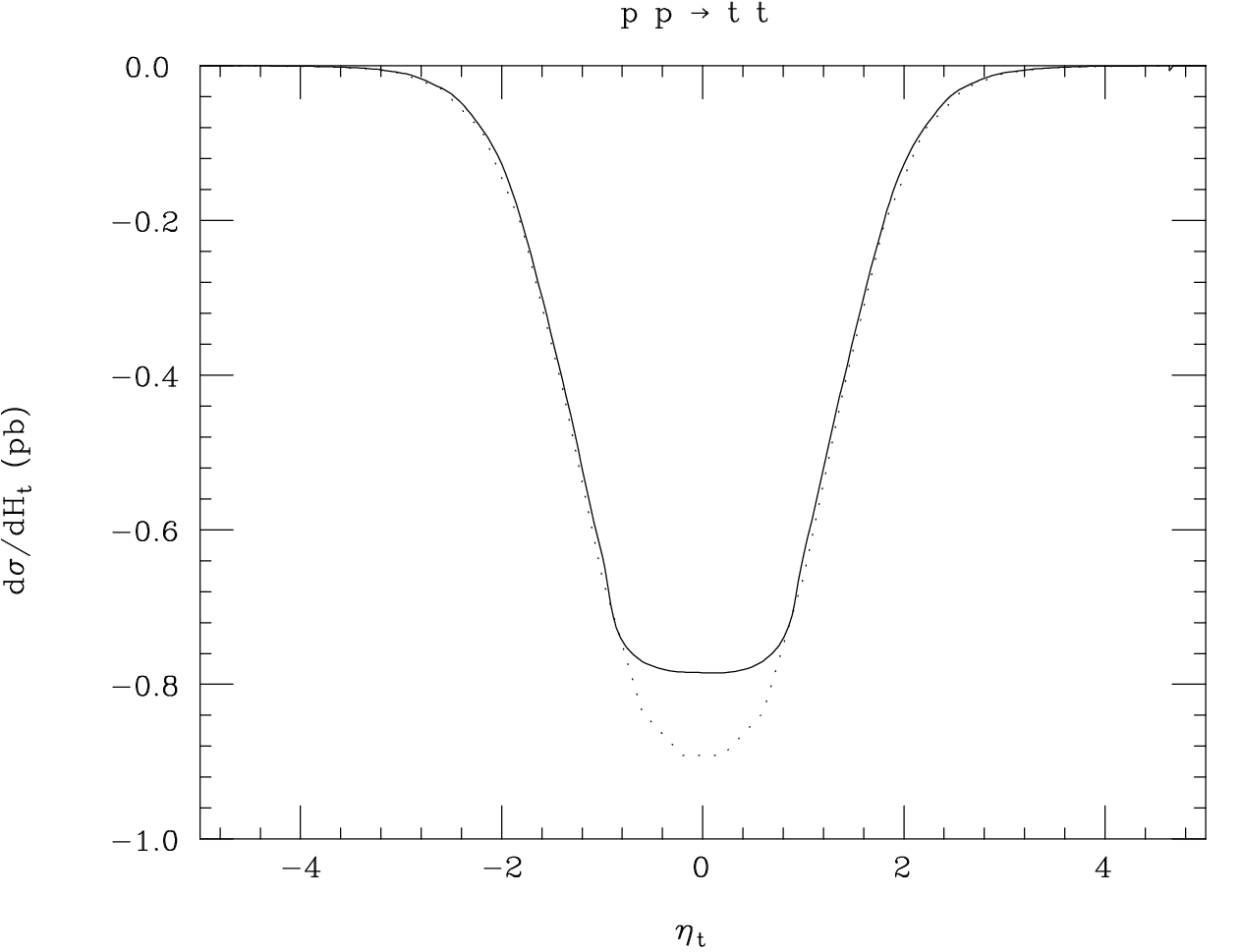}
\includegraphics[width=.375\linewidth,angle=90]{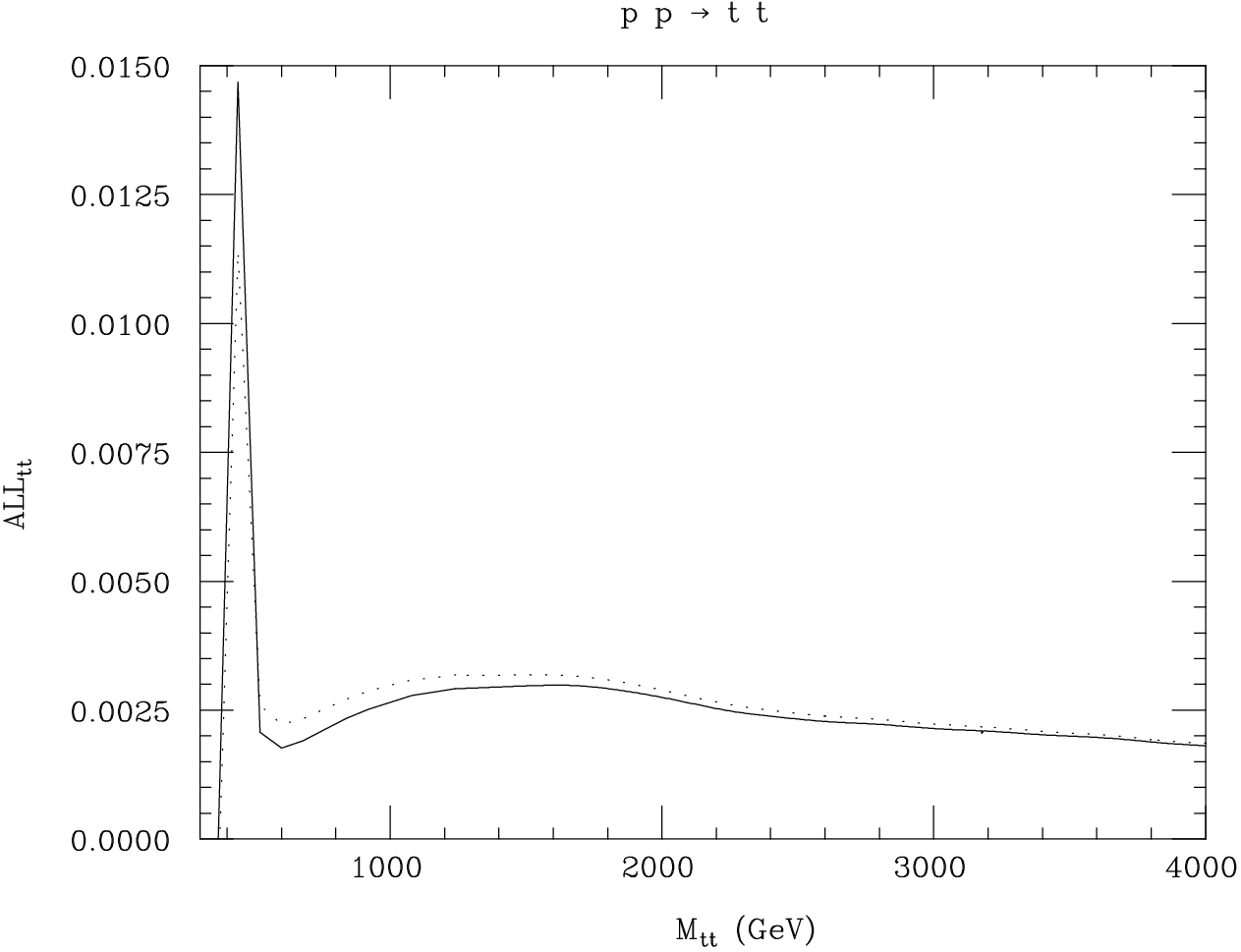}
\end{minipage}
\begin{minipage}{\textwidth}
\vspace{0.5cm}\hspace{-0.5cm}
\includegraphics[width=.375\linewidth,angle=90]{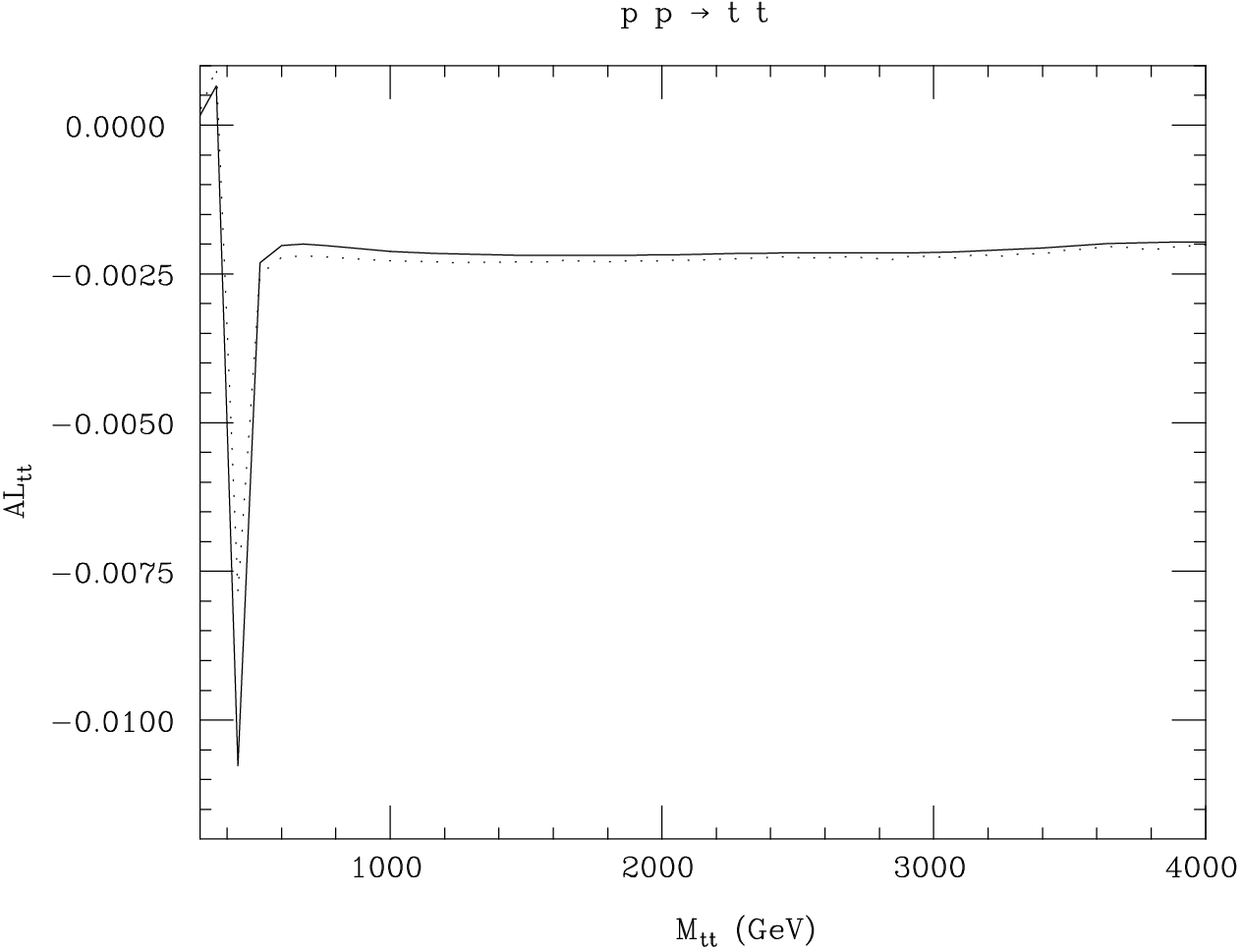}
\includegraphics[width=.375\linewidth,angle=90]{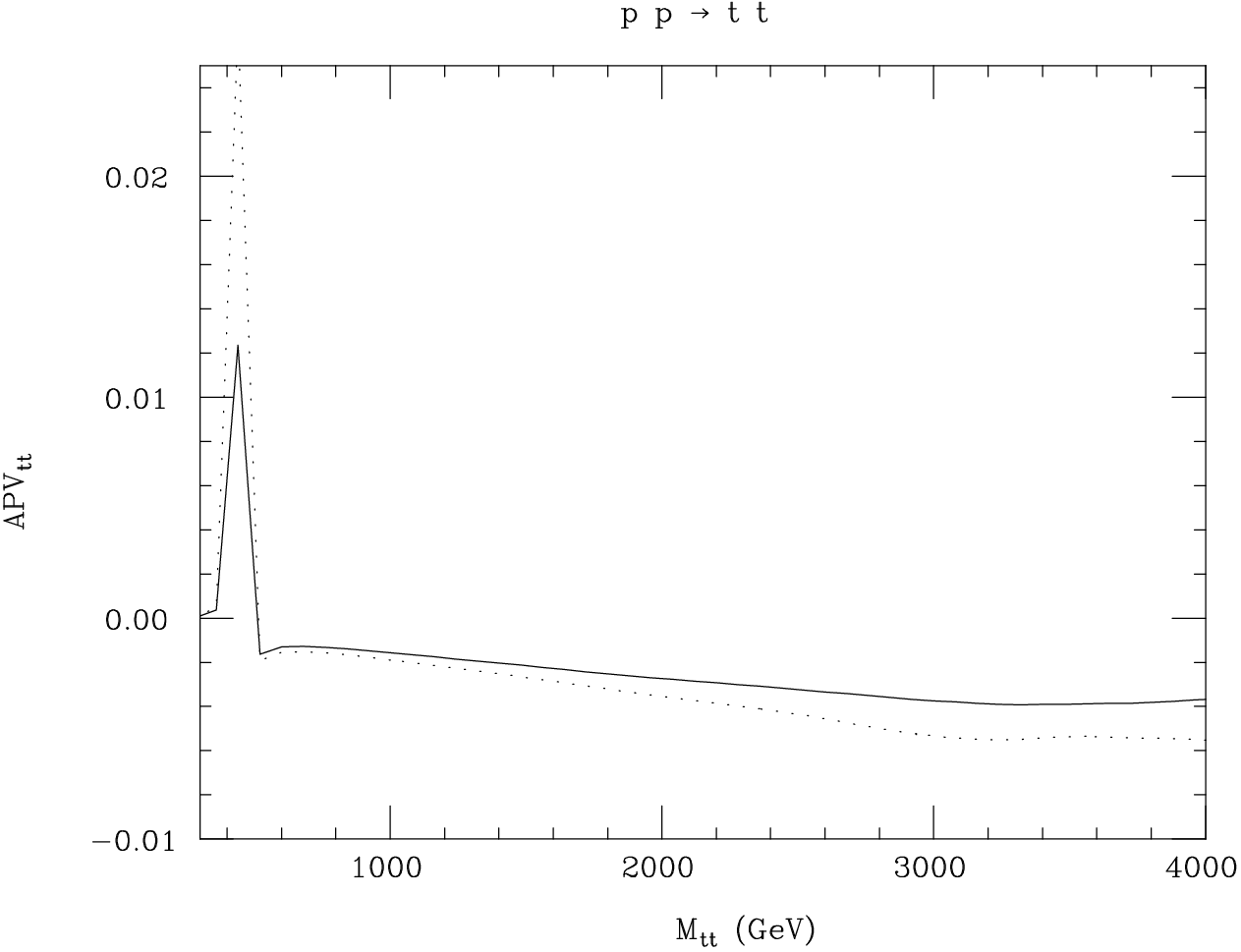}
\caption{The absolute size of the 
${\cal O}(\alpha_{\mathrm{S}}^2\alpha_{\mathrm{W}})$ 
corrections to the subprocess $gg\to t\bar t$ for the distribution in
(anti)top pseudorapidity $\eta_t$ 
(top-left frame) and the differential spin asymmetries 
(as defined in the text), for $M_H=150$ GeV (solid)
and $M_H=200$ GeV (dotted). The 
asymmetries are calculated along
the helicity axis as a function of the top-antitop invariant mass
$M_{t\bar t}$.}
\label{fig:Higgs}
\end{minipage}
\end{figure}

The dependence on the actual value of the Higgs mass is generally negligible
at both inclusive as well as differential level, with the possible 
exception of the pseudorapidity distribution in the very central region, see 
Fig.~\ref{fig:Higgs} (top-left frame), where the absolute size
of the ${\cal O}(\alpha_{\mathrm{S}}^2\alpha_{\mathrm{W}})$
corrections is shown for $M_H=150$ and 200 GeV. In fact,
the inclusive correction varies from $-2.33$ to $-2.51$ pb, respectively,
in comparison to a (Higgs independent) tree-level result of 384 pb.
For the other differential spectra studied such effects are rather uniformly
spread across the given kinematic range. As for the asymmetries, here
the effect of an increased Higgs mass varies significantly with
 $M_{t\bar t}$, yielding
differences with respect to the rates obtained with 
our default Higgs mass value which are not 
negligible over most of the kinematical
intervals considered, see Fig.~\ref{fig:Higgs}  (top-right and
bottom frames). 

\begin{figure}[!t]
\vspace*{1.0truecm}
\centerline{\hbox{\psfig{file=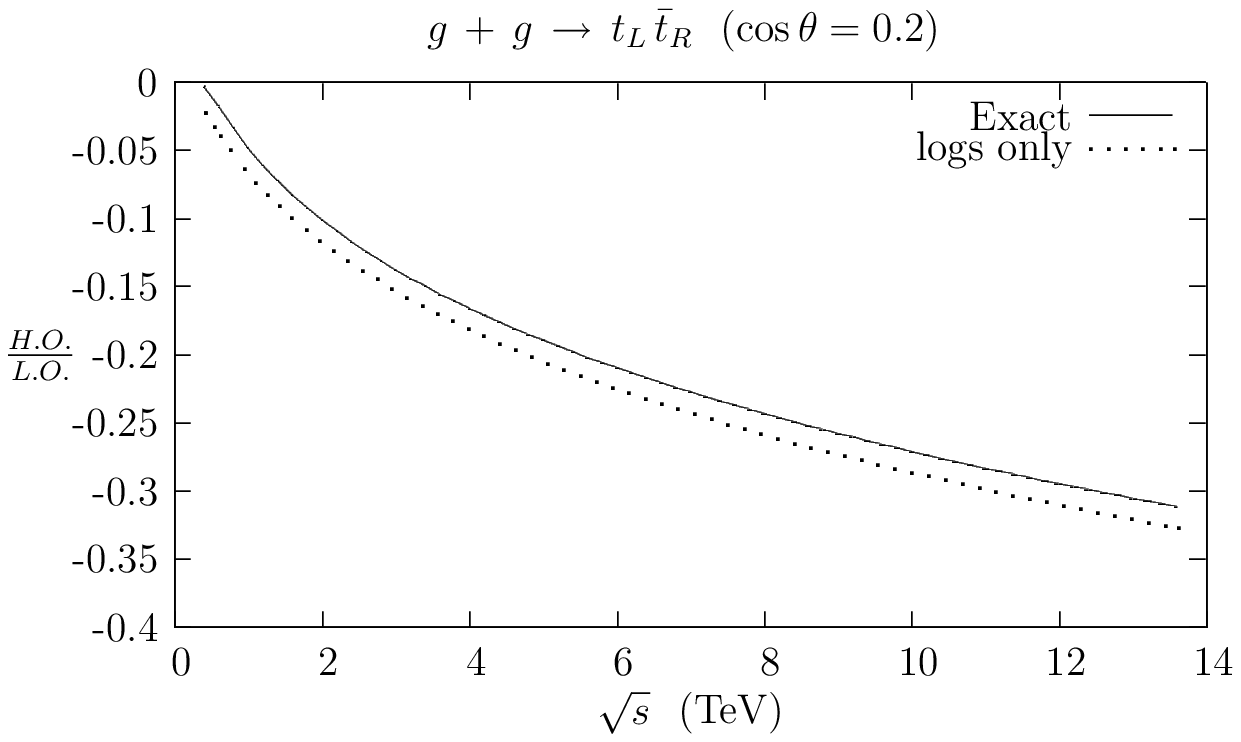,width=0.45\textwidth}}
            \hbox{\psfig{file=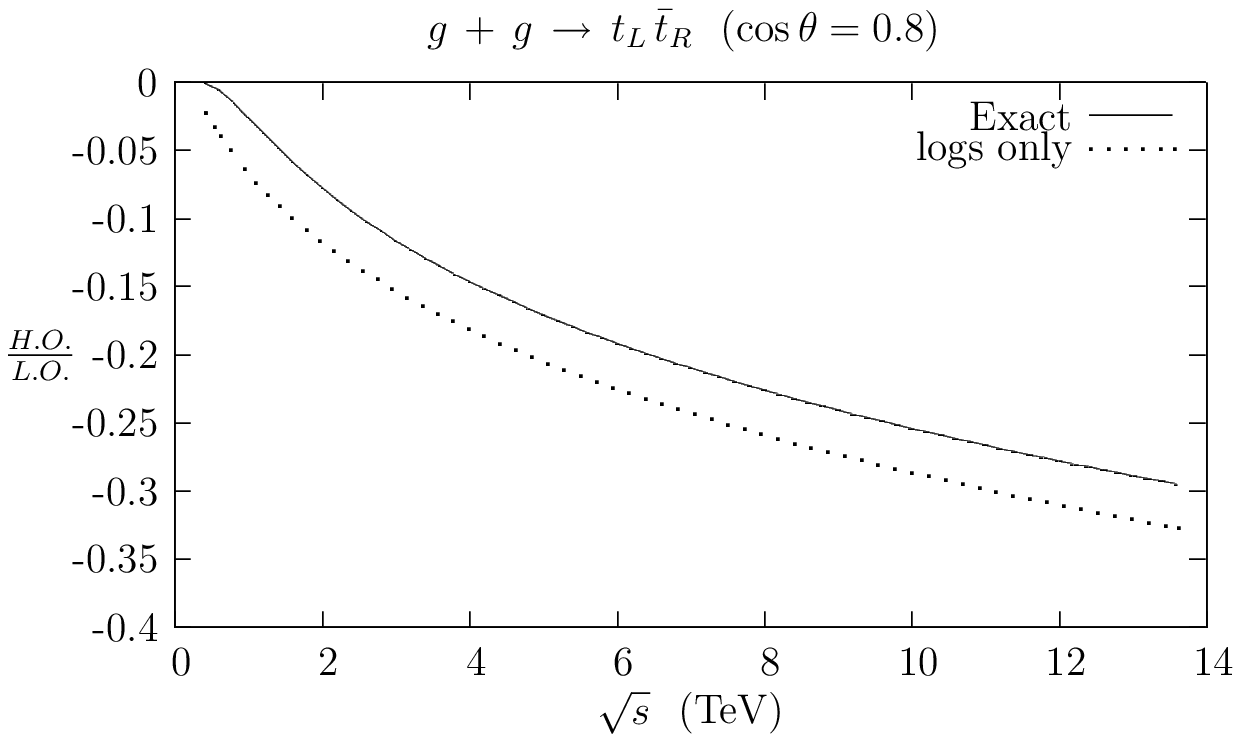,width=0.45\textwidth}}} 
\vspace*{1.0truecm}
\centerline{\hbox{\psfig{file=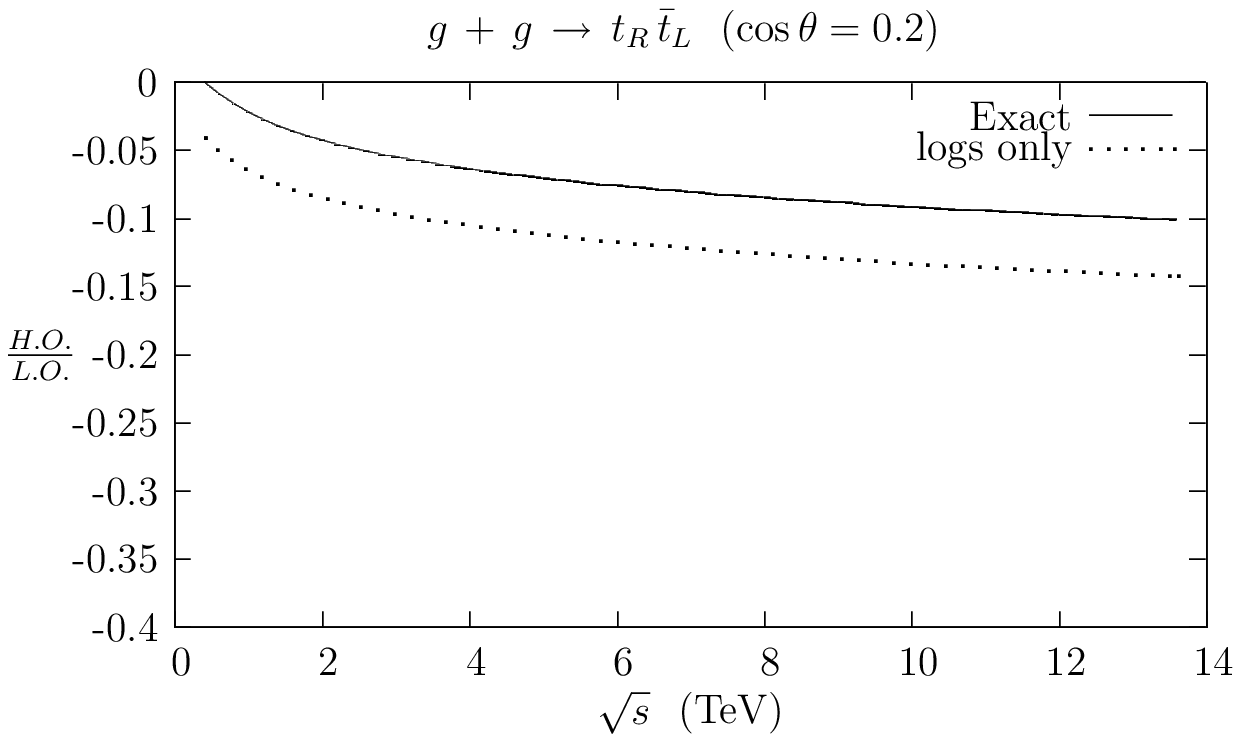,width=0.45\textwidth}}
            \hbox{\psfig{file=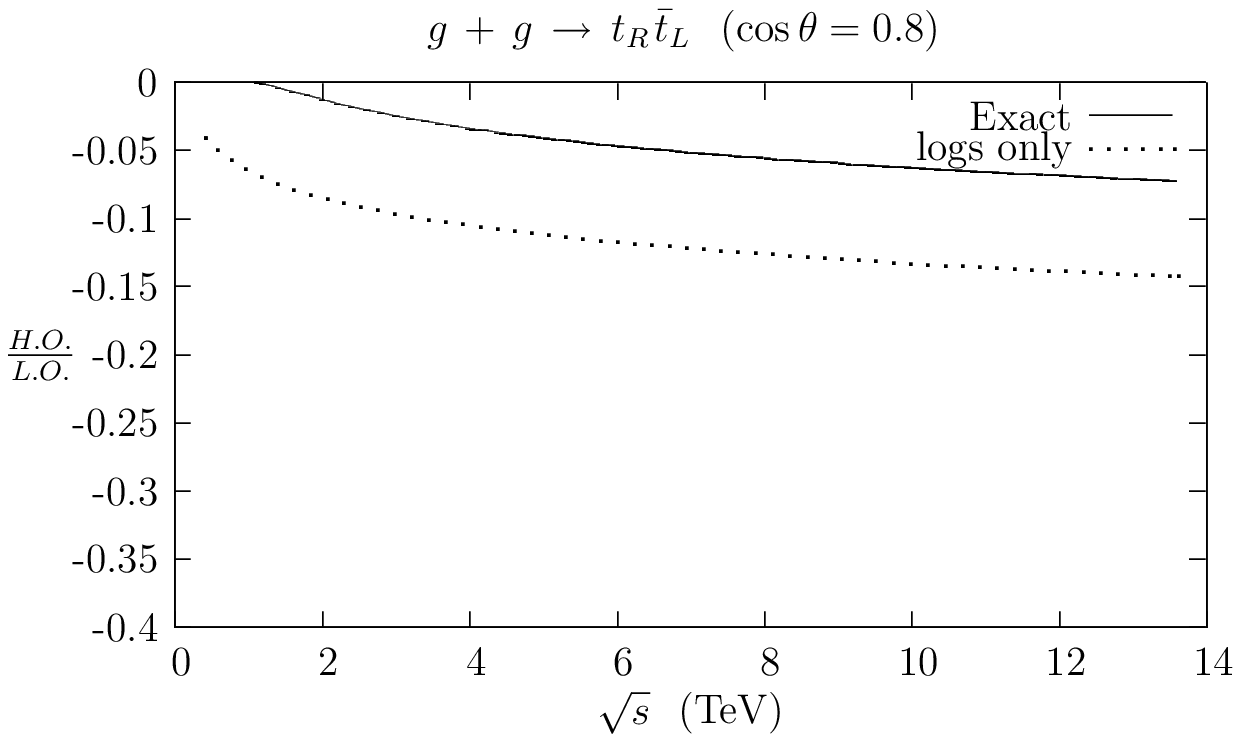,width=0.45\textwidth}}} 
\caption{A comparison of the exact ${\cal O}(\alpha_{\mathrm{S}}^2\alpha_{\mathrm{W}})$
corrections to those obtained from angular independent (double and 
single) logarithms only in Ref.~\cite{Marina}, 
for the two combinations with opposite helicities in the final state.
The graphs on the left(right) represent the case of large(small) angle
scattering. Notice that we have subtracted the EM contributions from
the formulae in Ref.~\cite{Marina}.}
\label{fig:verzegnassi}
\end{figure}

Before closing, it is of interest to compare our exact
${\cal O}(\alpha_{\mathrm{S}}^2\alpha_{\mathrm{W}})$
results with those of Ref.~\cite{Marina}. Notice that the latter only 
include the case of opposite helicities\footnote{The contribution
due to identical final state helicities becomes not negligible
near the threshold at $M_{t\bar t}\approx2m_t$.} 
in the final state and are limited
to the contribution of non-angular 
single ($\sim\alpha_{\rm{W}}\log  (s/M_W^2)$) and 
double ($\sim\alpha_{\rm{W}}\log^2(s/M_W^2)$) logarithms. 
In performing the comparison
between our results and those in  Ref.~\cite{Marina} we have removed the
EM component from the latter. From Fig.~\ref{fig:verzegnassi}, it is
clear that for the logarithmic approximation described be valid 
all Mandelstam variables $\hat s,\hat t,\hat u$ must be very large, 
condition which is 
obviously not fulfilled at small/large scattering angles. However, 
it should be appreciated that the ratio between the full
${\cal O}(\alpha_{\mathrm{S}}^2\alpha_{\mathrm{W}})$ term and the one obtained
in NLO Sudakov approximation is a constant to a very good approximation
already at moderate energies and for most angles.
Thus, in principle, a parameterisation of this constant as a function of 
the angle should be possible, in view of high statistics Monte Carlo 
simulations.

\section{Conclusions and outlook}

For kinematic variables accessible at the LHC, purely weak corrections through
one-loop level (without $Z$ bremsstrahlung) to the top-antitop cross section
via gluon-gluon fusion are generally small, although -- in order to
obtain both the appropriate normalisation and shape of the theoretical
prediction -- they cannot be neglected in the Sudakov
regime of some observables. In contrast, in line with the   
results reported in \cite{Maina:2003is}--\cite{PLB} for the
case of massless quark pair production, such 
one-loop weak effects are always
crucial in massive quark pair production when 
spin-asymmetries are considered (particularly, parity-violating 
ones). Ultimately, our results will have to be put
together with those from Refs.~\cite{Hans,Werner} (which we are in 
the process of repeating), in order to study
the top-antitop cross section at the level of precision required by the
LHC experiments. However, particular care
should eventually be devoted to the treatment of real $Z$ production and
decay in the definition of the inclusive data sample, as this will determine
whether (possibly positive) tree-level $Z$ bremsstrahlung effects have to be included
in the theoretical predictions through ${\cal O}(\alpha_{\mathrm{S}}^2\alpha_{\mathrm{W}})$,
which might counterbalance the negative effects due to the one-loop
$Z$ exchange estimated here. Along the same lines, it should be recalled that 
NNLO terms ought to be investigated too, as it is well known from the Sudakov
treatment that they may well be sizable in comparison to the NLO ones (see,
e.g., Ref.~\cite{twoloops}). Finally, we have verified that the effects 
studied here for the $gg\to t\bar t$ channel
are of no phenomenological relevance at the Tevatron. The experimental impact
of ${\cal O}(\alpha_{\mathrm{S}}^2\alpha_{\mathrm{W}})$ effects will eventually have to be assessed in a proper detector
simulation, in presence of top-antitop decay, parton shower and hadronisation: in fact, as shown
in \cite{ATLAS}, the possibility of extracting such effects is generally limited by systematics
rather than statistics. 

\section*{Acknowledgments}
We thank Claudio Verzegnassi for discussions and for financial support
during a visit to Trieste. SM also acknowledges useful conversations and
email exchanges with Matteo Beccaria, Paolo Ciafaloni and Denis Comelli.

\clearpage

\begin{flushright}
{SHEP-06-04E}\\
\today
\end{flushright}

\begin{center}
{\Large\bf Weak corrections to gluon-induced\\[0.15cm]
 top-antitop hadro-production: Erratum}
\vskip1.0cm
{\large S. Moretti, M.R. Nolten and D.A. Ross}\\
\vskip0.5cm
{\it School of Physics \& Astronomy, University of Southampton,
Southampton SO17 1BJ, UK} 
\end{center}

\centerline{\bf Abstract}
\vskip0.5cm\noindent
This is an Erratum to a paper of ours, 
Phys. Lett. B {\bf 639} (2006)  513. After its publication, 
we have discovered a mistake in a numerical program that
affects the results presented therein. We provide here the corrected version.

\section*{New numerical results}
\noindent
The numerical program used to produce the results of 
Ref.~\cite{published} was affected by a mistake, which has now been corrected.
We present in Figs.~1--3 the amended results, in correspondence to the
same figures in the original paper. We also note, as remarked in 
\cite{Werner0}, that the parity-violating asymmetry $A_{PV}$ defined in
\cite{published} is identically zero through 
${\cal O}(\alpha_{\mathrm{S}}^2\alpha_{\mathrm{W}})$, as we could
now verify explicitely. The inclusive results are now as follow:
while the (Higgs mass independent) tree-level cross section for
$gg\to t\bar t$ at the LHC is 384 pb, we now find that the weak corrections
through ${\cal O}(\alpha_{\mathrm{S}}^2\alpha_{\mathrm{W}})$ amount
to  $-9.65$ pb for $M_H=150$ GeV and   
    $-9.39$ pb for $M_H=200$ GeV.  All these plots and cross sections correspond
to the numerical setup (masses, couplings, PDFs, scales, etc.) declared
in \cite{published}.

For all such results, as well additional ones presented in 
Refs.~\cite{Werner0,Hans0} (where independent calculations of the same corrections tackled
in \cite{published} were performed), we found very good agreement between ourselves
and Refs.~\cite{Werner0,Hans0}, for the same choice of input parameters. 

As for Fig.~4 of \cite{published}, here the differences are not substantial.
Besides, the purpose of those plots was to illustrate the difference between
the full results and the Sudakov approximation, which is largely unaffected
by the correction we made to our program. Hence, we do not reproduce those
plots here. 

Finally, we have now also calculated the ${\cal O}(\alpha_{\mathrm{S}}^2\alpha_{\mathrm{W}})$
corrections to the $q\bar q\to t\bar t$ subprocess and the results obtained (not shown here)
are in agreement with the corresponding ones in Refs.~\cite{Werner1,Hans1}.

\section*{Acknowledgments}
We are extremely grateful to Werner Bernreuther, Michael F\"ucker and
Peter Uwer for their help in checking our program and finding the source
of the discrepancies between our results and theirs.

\begin{figure}[!t]
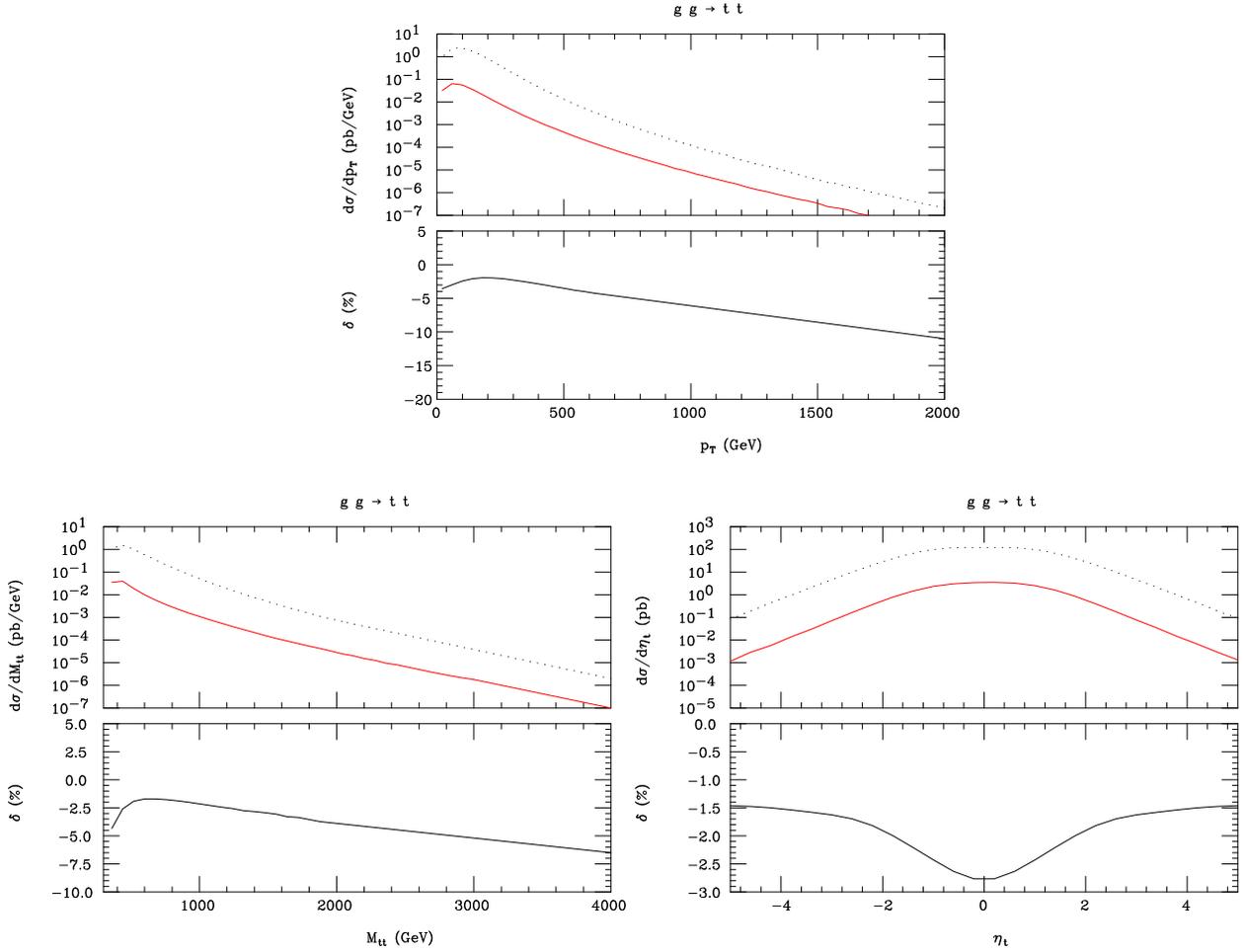

\begin{minipage}{\textwidth}
\hspace{3.95cm}
\includegraphics[width=.3725\linewidth,angle=90]{Erratum_pT.ps}
\end{minipage}
\begin{minipage}{\textwidth}
\vspace{0.5cm}\hspace{-0.5cm}
\includegraphics[width=.3725\linewidth,angle=90]{Erratum_M.ps}
\includegraphics[width=.3725\linewidth,angle=90]{Erratum_eta.ps}
\caption{Differential distributions of the subprocess
$gg\to t\bar t$ through the ${\cal O}(\alpha_{\mathrm{S}}^2)$
(top frames, dotted) and the 
${\cal O}(\alpha_{\mathrm{S}}^2\alpha_{\mathrm{W}})$ 
(top frames, solid) 
as well as the percentage of the latter with respect to the former
(bottom frames, solid)  
for the (anti)top transverse momentum $p_T$, the top-antitop invariant mass
$M_{t\bar t}$ and the (anti)top pseudorapidity $\eta_t$. 
(Lightly/Red coloured solid tracts in logarithmic scale 
are intended to be negative.)}
\end{minipage}
\end{figure}

\clearpage

\begin{figure}[!t]
\begin{minipage}{\textwidth}
\hspace{-0.5cm}
\includegraphics[width=.3725\linewidth,angle=90]{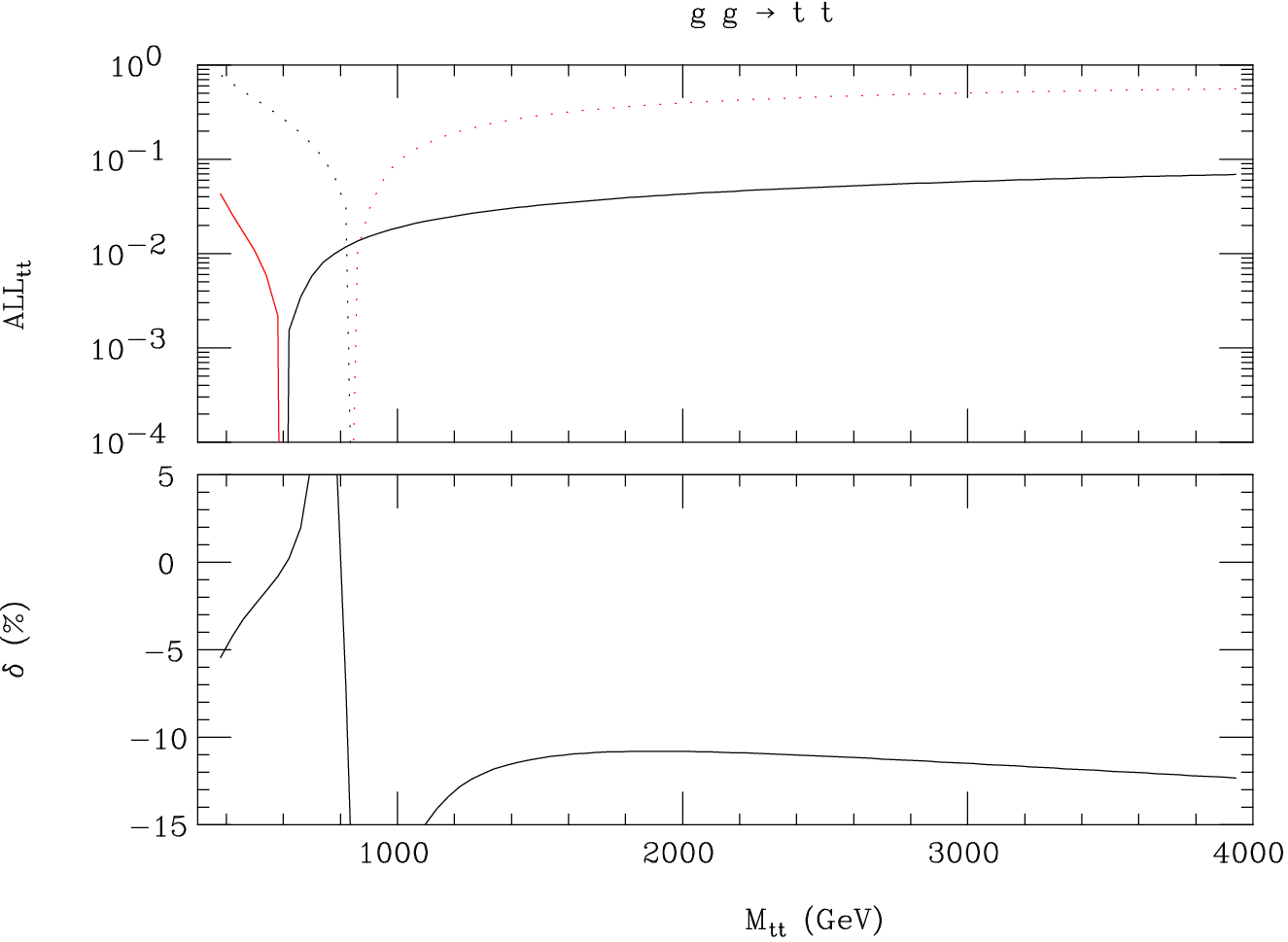}
\includegraphics[width=.3725\linewidth,angle=90]{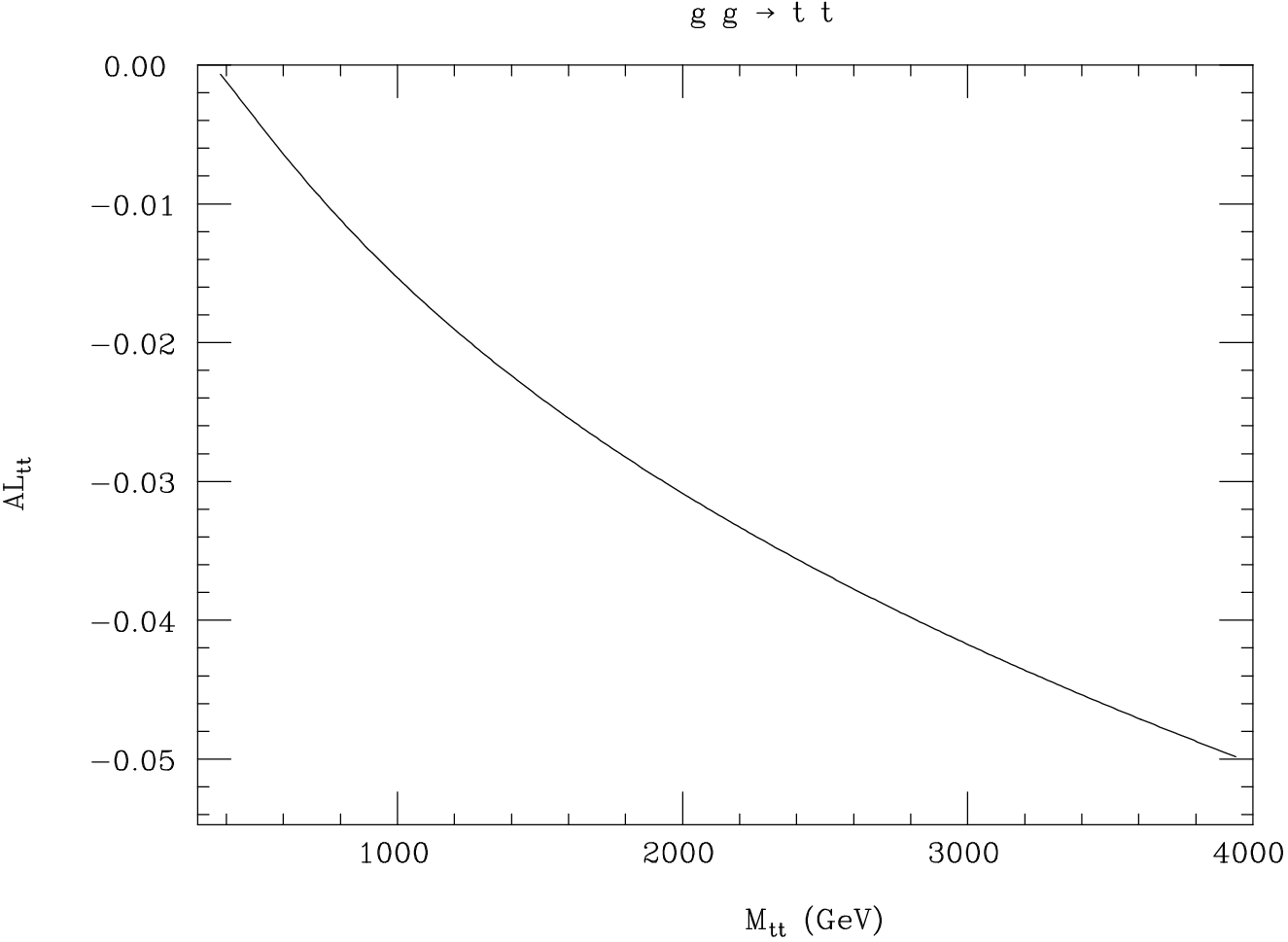}
\caption{The differential spin asymmetry $A_{LL}$ (as defined in 
 \cite{published})
of the subprocess
$gg\to t\bar t$ through the ${\cal O}(\alpha_{\mathrm{S}}^2)$
(left plot, top frame, dotted) and the 
${\cal O}(\alpha_{\mathrm{S}}^2\alpha_{\mathrm{W}})$ 
(left plot, top frame, solid).
(Note that the LO QCD contribution changes sign
at $\approx900$ GeV and is heavily dependent on $M_{t\bar{t}}$ whereas the 
${\cal O}(\alpha_{\mathrm{S}}^2\alpha_{\mathrm{W}})$ 
correction is not.) In the left plot, bottom frame, we show the 
percentage correction
to the (non-zero) LO QCD asymmetry for $A_{LL}$ due to 
${\cal O}(\alpha_{\mathrm{S}}^2\alpha_{\mathrm{W}})$  effects.
The right plot displays the asymmetry $A_L$ 
(as defined in \cite{published}), which vanishes
exactly in LO QCD, through the same order.
The 
asymmetries are calculated along
the helicity axis as a function of the top-antitop invariant mass
$M_{t\bar t}$. 
}
\end{minipage}
\end{figure}

\clearpage

\begin{figure}[!t]
\begin{minipage}{\textwidth}
\hspace{3.5cm}
\includegraphics[width=.3725\linewidth,angle=90]{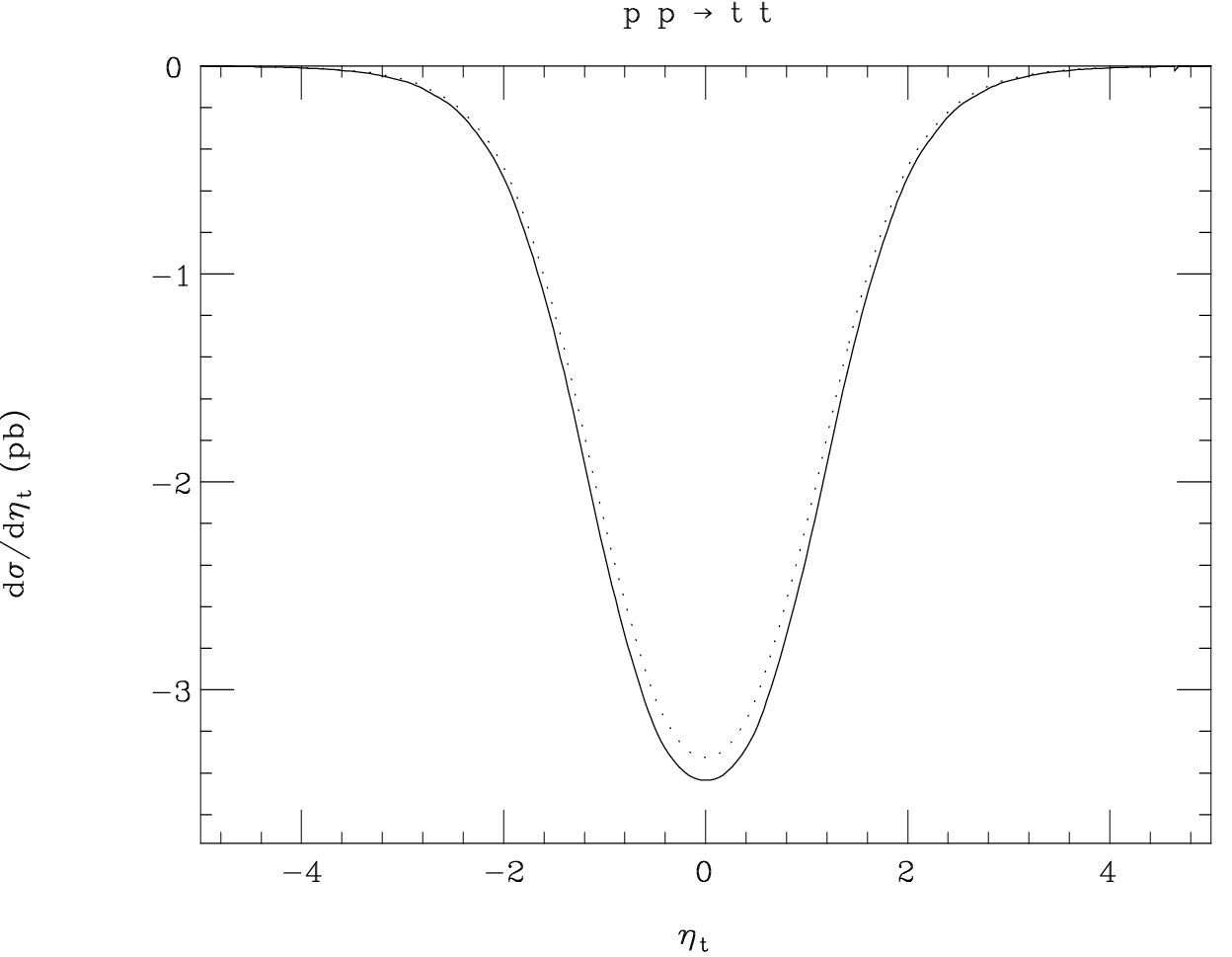}
\end{minipage}
\vspace{0.5cm}
\begin{minipage}{\textwidth}
\hspace{-0.5cm}
\includegraphics[width=.3725\linewidth,angle=90]{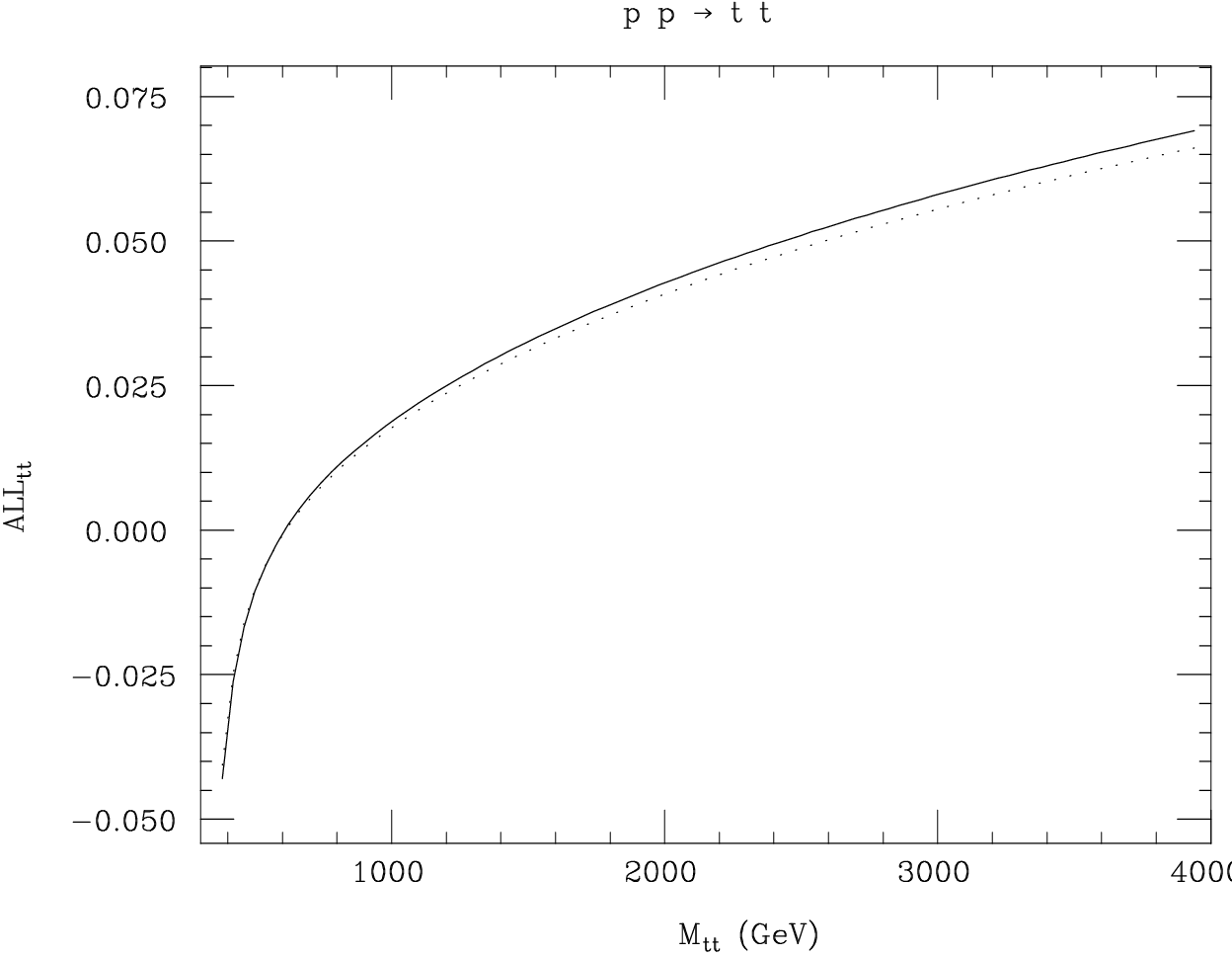}
\includegraphics[width=.3725\linewidth,angle=90]{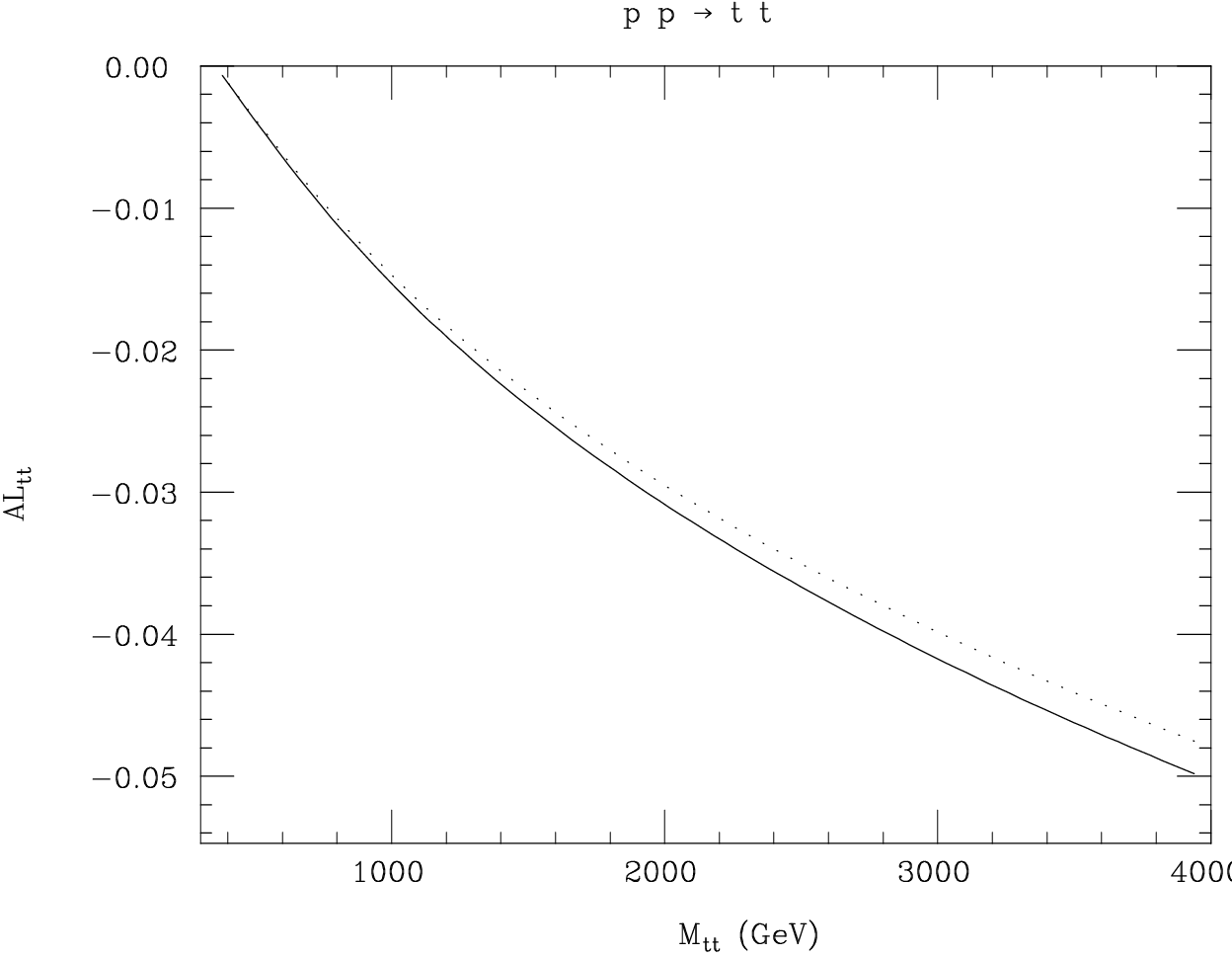}
\caption{The absolute size of the 
${\cal O}(\alpha_{\mathrm{S}}^2\alpha_{\mathrm{W}})$ 
corrections to the subprocess $gg\to t\bar t$ for the distribution in
(anti)top pseudorapidity $\eta_t$ 
(top plot) and the differential spin asymmetries 
(bottom plots, as defined in \cite{published}), for $M_H=150$ GeV (solid)
and $M_H=200$ GeV (dotted). The 
asymmetries are calculated along
the helicity axis as a function of the top-antitop invariant mass
$M_{t\bar t}$.}
\end{minipage}
\end{figure}


\end{document}